\documentclass[twocolumn,showpacs,pra,eqsecnum,citeautoscript,amsmath,amssymb,floatfix,groupedaddress]{revtex4-1}
\usepackage{bm,xcolor,amsmath,amssymb,mathrsfs,latexsym,graphicx,psfrag,float}
\usepackage{dcolumn}
\usepackage{hyperref}

\usepackage{leftidx}

\usepackage{tabularx}

\usepackage[english]{babel}
\usepackage[utf8]{inputenc}
\usepackage{bbold}

\usepackage{tikz}
\usetikzlibrary{arrows,decorations.pathmorphing,backgrounds,positioning,fit,matrix}
\usepackage{natbib}

\usepackage{hyperref}
\usepackage{xcolor}
\usepackage{cleveref} 
\usepackage{graphicx}
\graphicspath{{pics/}} 
\usepackage{wrapfig}
\usepackage[caption=false]{subfig}
\usepackage{subfloat}
\usepackage{isotope}
\usepackage{dcolumn}
\usepackage{color}
\usepackage{placeins}
\DeclareGraphicsExtensions{.pdf,.png,.jpg}

\usepackage{amsmath}
\usepackage{bbm}
\usepackage{bbold}
\usepackage{empheq} 
\usepackage{braket}

\definecolor{dark-red}{rgb}{0.4,0.15,0.15}
\definecolor{dark-blue}{rgb}{0.15,0.15,0.4}
\definecolor{medium-blue}{rgb}{0,0,0.5}
\hypersetup{
colorlinks, linkcolor={dark-blue},
citecolor={dark-blue}, urlcolor={medium-blue}
}

\renewcommand{\l}{\left(}
\renewcommand{\r}{\right)}

\renewcommand{\b}[1]{\boldsymbol{#1}}

\newcommand\scalemath[2]{\scalebox{#1}{\mbox{\ensuremath{\displaystyle #2}}}} 

\makeatletter
\newcommand{\subalign}[1]{%
  \vcenter{%
    \Let@ \restore@math@cr \default@tag
    \baselineskip\fontdimen10 \scriptfont\tw@
    \advance\baselineskip\fontdimen12 \scriptfont\tw@
    \lineskip\thr@@\fontdimen8 \scriptfont\thr@@
    \lineskiplimit\lineskip
    \ialign{\hfil$\m@th\scriptstyle##$&$\m@th\scriptstyle{}##$\crcr
      #1\crcr
    }%
  }
}
\makeatother

\begin{document}

\title{\texorpdfstring{Many-Body Quantum Optics with Decaying Atomic Spin States:\\
($\gamma$, $\kappa$) Dicke model}{Many-Body Quantum Optics with Decaying Atomic Spin States:\\
(gamma, kappa) Dicke model}}

\author{Jan Gelhausen}
\author{Michael Buchhold} 
\author{Philipp Strack}

\affiliation{Institut f\"ur Theoretische Physik, Universit\"at zu K\"oln, D-50937 Cologne, Germany}

\date{\today}

\begin{abstract}
We provide a theory for quantum-optical realizations of the 
open Dicke model with internal, atomic spin states subject to spontaneous emission with rate $\gamma$.
This introduces a second decay channel for 
excitations to irreversibly dissipate into the environment, in addition to
the photon loss with rate $\kappa$, which is composed of individual atomic decay processes and a collective atomic decay mechanism. The strength of the latter is determined by the cavity geometry.
We compute the mean-field non-equilibrium steady states for spin and photon 
observables  in the 
long-time limit, $t\rightarrow \infty$. 
Although $\gamma$ does not conserve 
the total angular momentum of the spin array, we argue that our solution is 
exact in the thermodynamic limit, for the
number of atoms $N\rightarrow \infty$.
In light of recent and upcoming 
experiments realizing superradiant phase transitions using internal atomic states 
with pinned atoms in optical lattices, our work lays the foundation for the
pursuit of a new class of open quantum magnets coupled to quantum light.
\end{abstract}
\maketitle

\section{Introduction}
Significant research efforts in the science 
of quantum optics are directed towards ``scaling up''
the minimal building block of one atomic qubit and 
one single photon towards $N$ qubits and $M$ photon modes.
One objective is that a controllable 
assembly of such systems in the quantum regime 
has all the ingredients of a quantum computer including 
channels for communication of information within 
quantum networks \cite{gardiner15}.
This development represents both, a tremendous opportunity 
and a challenge, to seriously study the many-body physics of 
extended quantum-optical systems, where arrays of 
qubits are coherently coupled to quantum light.

To that end, a promising recent experimental development is the realization 
of tunable lattice potentials within resonators hosting photons with 
optical wavelengths \cite{klinder15,landig16,neuzner16}.
These set-ups allow controllable placement of large numbers of quantum emitters, 
in the form of ultracold atoms, into lattice sites, preserving their relative 
phases. The atom-cavity coupling now has ``single-site resolution'' by 
overlaying cavity mode functions with lattice potentials and the targeted 
loading process of atoms into given lattice sites.

Making the lattice potentials sufficiently deep, one can now 
access a regime in which the atomic motion is suppressed 
completely and the dynamics of \emph{internal, spin excitations} 
play the lead role. Although the analogy is dangerous and incomplete, 
let us mention that a corresponding situation in an electronic condensed matter 
material would be a Mott insulator, in which the charge degrees of freedom 
are localized, and the electronic spins interact via, typically short-ranged, 
exchange couplings $J_{ij}$.  As was recognized theoretically
a few years back, the cavity set-up allows a much richer set 
of $J_{\ell m}$'s (variable range, complex- vs. real-valued) and 
unconventional magnetic phases to be realized \cite{strack11,sarang11,bunti13}.
However, the ``drosophila'' in this field is the Dicke model, an infinite-ranged, exactly 
solvable ferromagnet \cite{hepp73,wang73}, which has recently also been realized 
experimentally using internal spin states \cite{dimer07,baden14}.

A basic physical difference to the earlier realizations of the 
Dicke model using \emph{momentum states} of a thermal or condensed Bose gas
\cite{black03,baumann10} is the increased fragility of internal, spin states 
to dissipative processes such as atomic spontaneous emission. Indeed, the decay rate of 
collective momentum modes $\gamma_{\rm mom}$
of an atomic gas is remarkably small ($\gamma_{\rm mom} \ll \kappa  \lesssim g $) 
and limited mostly by thermal effects and collisions 
\cite{brennecke13, piazza13,kulkarni13,konya14,piazza14}.
By contrast single-site atomic spontaneous emission with rate $\gamma$ 
tends to deplete the system of excitations and driving each spin into the $|\downarrow \rangle$ 
state. There is no analog of this dissipative process for momentum states and therefore 
its basic physical effects have not been explored much in this context. Moreover, 
the experiments by Baden {\it{et al.}} \cite{baden14} were not entirely able to compare their data to 
a theory for the open Dicke model with spontaneous emission, clearly identifying 
a gap in the current literature.

The objective of this paper is to reveal the interplay of spontaneous emission 
with the collective interactions induced by the resonator. We extend previous 
works of the open Dicke model \cite{dimer07}, which were restricted to photon 
losses, to the full two loss channels $(\gamma,\kappa)$ variant. The atomic spontaneous emission consists of single-site atomic decay and a collective atomic contribution \cite{Scully2015} whose strength is controlled by the cavity geometry in a large sample limit. For the lattice based experiments considered in this work, the collective atomic decay is much smaller than all other typical scales, which we demonstrate below, and we restrict the present analysis to the limit of weak collective decay.
By this, we want to lay the foundation for the study of interacting, open quantum magnets
with atoms in optical lattices in many-body cavity QED 
\cite{klinder15,landig16,neuzner16} and other nano-photonic setups 
such as atoms trapped close to photonic crystals \cite{thompson12,chang13}.

\newpage
\subsection{Key results and outline of paper}

Our main result is the derivation of the exact formula of the critical coupling for the onset 
of superradiance in Sec.~\ref{subsec:critical} in the presence 
of both single-site and collective atomic spontaneous emission. In Sec.~\ref{subsec:singapore}, we use this result 
to address an observed discrepancy between experimental data for the 
critical pump strength and earlier calculations. The argument why this 
formula remains exact in the presence of atomic spontaneous emission $\gamma$, is given in Sec.~\ref{subsec:exact}. In short, site-to-site 
variances between observables vanish in the thermodynamic limit, because 
the Hamiltonian affects only the homogeneous, zero-momentum component of the 
spins. It is true that $\gamma$ also couples to finite-$k$ components, in constrast 
to the $\kappa$-only Dicke model, but $\gamma$ just leads to their decay. In Sec.~\ref{subsec:singapore}, we compare the prediction with the onset of superradiance with experimental data of the Singapore group. There we also discuss and compare different decay mechanisms and rates of atomic excitations.

Based on the Heisenberg-Langevin equations derived in Sec.~\ref{subsec:langevin}, 
we compute the values of non-equilibrium steady states, 
the cavity output spectrum, and the effective temperature of the photons at the superradiance transition
in Sec.~\ref{subsec:steady}-\ref{subsec:Teff}.

\section{\texorpdfstring{$(\gamma,\kappa)$ Dicke model}{(gamma,kappa) Dicke model}}
\label{sec:model}
In this section, we begin by explaining the model Dicke Hamiltonian 
and the Liouvillians for the two decay processes: photon loss 
and atomic spontaneous emission.
Then, we connect this model to a recent quantum optics experiment, wherein the spin states 
in the Dicke Hamiltonian were realized via two atomic hyperfine-split levels.
We finally present the Heisenberg-Langevin and the mean-field master equations within the Markov approximation.

\subsection{Hamiltonian and Liouvillians}
\label{subsec:hamiltonian}
The core of the set-up is an array of $N$ atomic spins at 
fixed positions in space that interact with a single cavity mode. The Dicke-Hamiltonian for the system describes coherent exchange of atomic excitations with excitations of the light field, 
\begin{align}
H&=\omega_0 a^{\dagger}a+\left(\frac{U}{2N}\sum_{\ell=1}^{N}\sigma^z_{\ell}\right)a^{\dagger}a
+\frac{\Delta}{2}\sum_{\ell=1}^{N}\sigma^z_{\ell}\nonumber \\
&+\frac{g}{\sqrt{N}}(a+a^{\dagger})\sum_{\ell=1}^{N}(\sigma^+_{\ell}+\sigma^-_{\ell})\;.
\label{eq:hamiltonian}
\end{align}
Here there is also an (effective) photon energy $\omega_0$, a longitudinal field in 
$z$-direction for the spins $\Delta$, and an additional frequency shift of the photons due to 
a coupling $U$ to the collective $z$-component. This last coupling 
arises in the quantum-optical implementation we discuss below in 
Subsec.~\ref{subsec:exp}.

The second class of processes introduce decoherence and are 
irreversible decay processes of both, photonic (rate $\kappa$) and atomic excitations 
(rate $\gamma$) into the reservoir modes of the electromagnetic vacuum surrounding 
the cavity. The photons decay through the imperfect mirrors and the atoms directly decay into the reservoir modes via the solid angle not covered by the mirrors, as shown in Fig.\,\ref{Fig:ExperimentalSetUp}. Their effect can be captured by introducing the 
Lindblad operators, which act on the system density matrix $\rho$
in the following way:
\begin{align}
\mathcal{L}_\gamma[\rho]=&\gamma(1-\alpha)\sum_{\ell=1}^N \left( \sigma^{-}_{\ell}\rho\sigma^{+}_{\ell}-\frac{1}{2}\{\sigma^{+}_{\ell}\sigma^{-}_{\ell},\rho\}\right)\label{EQ:LindbladOperatorgamma} \\
&+\gamma \alpha \left( S^{-}\rho S^{+}-\frac{1}{2}\{S^{+}S^{-},\rho\}\right),\nonumber \\
\mathcal{L}_\kappa[\rho]&=\kappa\bigg[2 a \rho a^{\dagger}-\{a^{\dagger}a ,\rho\}\bigg]\;.
\label{eq:lindblads}
\end{align}
Here, the atomic spontaneous emission consists of two contributions. The first is the single atom decay rate $\gamma$ for a single atom at site $\ell$. The second contribution is a collective decay contribution expressed with the collective atomic operators $S^{\pm}=\sum_{\ell=1}^N \sigma^{\pm}_{\ell}$. The prefactor that controls the strength of the collective decay is bounded, $0\le\alpha\le1$, and depends on the cavity geometry. While a detailed derivation of the atomic Lindblad operator can be found in App.~\ref{App:SecCollectiveAtomicDecay}, we want to stress here that the lattice setups under consideration generally correspond to $\alpha\ll 1$. We also demonstrate that the collective contribution to the decay correctly reproduces super- and subradiant decay rates for collective atomic states in App.\,\ref{App:SubandSuperradiantDecay} .  
It can be seen that in such a description the single atom loss term scales linearly in the atom number whereas the collective loss term scales quadratically with the atom number. Since every experimental system is necessarily finite with a well-defined atom-number $N_0$, so is the collective loss rate. In order to define a sensible thermodynamic limit, for which both the average energy and loss rate per particle remain constant, the geometric coupling term is rewritten as $\gamma \alpha \to \gamma\alpha N/N=\beta / N$. The thermodynamic limit is now understood as taking $N\to \infty$ and $V \to \infty$ with $N/V=const.$ and $\beta=\alpha N_0=const.$, where $N_0$ is the experimentally relevant number of atoms and therefore fixes the collective loss rate. This is analogous to the thermodynamic limit of the Dicke Hamiltonian, for which the coupling of the light field to an individual atom is finite and fixed in any experimental set-up. However, the correct description of the system in the thermodynamic limit necessitates that the coupling is written as $\sim\frac{g}{\sqrt{N}}\sum_{\ell=1}^N\sigma^{x}_{\ell}(a+a^{\dagger})$ such that in the thermodynamic limit $g$ is constant and fixed.

The reservoirs have Markovian character; 
this is because the Hamiltonian Eq.~(\ref{eq:hamiltonian}) becomes time-independent 
only in a frame rotating with an optical (pump) frequency. In this frame, bath and system time scales are 
well separated by orders of magnitude. 

The interplay and competition between unitary and irreversible dynamics
can be studied with a Master equation for the density matrix
\begin{align}
\dot\rho &=-i[H,\rho]+\mathcal{L}_{\kappa}[\rho]+\mathcal{L}_{\gamma}[\rho]\;.
\label{eq:master}
\end{align}
As the Hamiltonian in Eq.\,\eqref{eq:hamiltonian} does not conserve the total number of excitations $\mathcal{N}=a^{\dagger}a+\frac{1}{2}\sum_{\ell=1}^N \sigma^z_{\ell}+\frac{N}{2}$ the Hamiltonian will counteract the depletion processes of the Lindblad terms. This is in contrast to a Hamiltonian where the counter-rotating terms are dropped in a rotating-wave-approximation and for which there would be no other steady-state than the empty dark state. 

\subsection{\texorpdfstring{Exact solvability in long-time limit $t\rightarrow \infty$ and thermodynamic 
limit $N\rightarrow \infty$}{Exact solvability in long-time limit t to infty and thermodynamic 
limit N to infty}}
\label{subsec:exact}

It is known that the Dicke model in thermodynamic equilibrium is exactly solvable by a mean-field ansatz 
\cite{hepp73,wang73}. Although the exact solutions get more complicated, 
this remains true for the non-equilibrium steady states of atomic quantum gases 
in optical cavities \cite{piazza13,piazza14} provided one takes 
first the thermodynamic limit, number of atoms $N\rightarrow \infty$, 
and then the long time limit, $t\rightarrow \infty$ ("$t-N$ limit").
Now one may wonder whether this remains true in the presence of single and collective loss rates in the atomic sector. 

Here, we present a brief argument, which shows that mean-field 
non-equilibrium steady states solve 
Eqs.~(\ref{eq:hamiltonian}-\ref{eq:master})
exactly in the $t-N$ limit. To see that, we first integrate 
out the photons. This can be done exactly 
retaining photon losses and other pump and loss terms 
for the photons as long as they are quadratic in the photon fields
\cite{dalla13}.
 This yields a ferromagnetic all-to-all coupling
 $-J/N \left(\sum_{\ell = 1}^N \sigma_\ell^x\right)
 \left( \sum_{m=1}^N\sigma^x_m\right)$ mediated by photon exchange. 
 We may set $U=0$ in Eq.~(\ref{eq:hamiltonian}).
We now go the thermodynamic limit $N\rightarrow \infty$ and write 
the Hamiltonian in momentum-space. Momenta $k$
are now continuous variables and integrated over $\int_k$ 
the appropriate Brillouin zones
\begin{align}
H_{\rm eff} &= \frac{\Delta}{2}
\int_k \delta_{k,0}\, \sigma^z_k
-
J \int_k  \delta_{k,0} \,
\sigma^x_k \sigma^x_{-k}\;.
\label{eq:Hamiltonian_Fourier}
\end{align}
The remaining Lindblad operator for $\gamma$ reads in momentum space
\begin{align}
\mathcal{L}_\gamma[\rho]=&\gamma(1-\alpha)\int_k
\bigg[\sigma^{-}_{k} \rho \sigma^{+}_{k}-\frac{1}{2}\{\sigma^{+}_{k}\sigma^{-}_{k},\rho\}\bigg]\\
&+\gamma \alpha N_0\int_{k,k'}\delta_{k,0} \delta_{k',0} \bigg[\sigma^{-}_{k} \rho \sigma^{+}_{k'}-\frac{1}{2}\{\sigma^{+}_{k}\sigma^{-}_{k'},\rho\}\bigg]
\nonumber\\
=& 
\gamma(1+\beta) \int_k
\delta_{k,0}
\bigg[\sigma^{-}_{k} \rho \sigma^{+}_{k}-\frac{1}{2}\{\sigma^{+}_{k}\sigma^{-}_{k},\rho\}\bigg]
\nonumber \\&+
\gamma\int_{k\neq 0}
\bigg[ \sigma^{-}_{k} \rho \sigma^{+}_{k}-\frac{1}{2}\{\sigma^{+}_{k}\sigma^{-}_{k},\rho\}\bigg]\;,
\label{EQ:Lindblad_Fourier}
\end{align}
where in the second line we have split off the decay for the 
zero-momentum component from the finite momentum components and have set $\beta=\alpha (N_0-1)$ as the fixed coupling strength in the thermodynamic limit.
The Hamiltonian dynamics Eq.~(\ref{eq:Hamiltonian_Fourier}) 
operates strictly within only the zero-momentum sub-space of the spins 
(in position space this is the homogeneous component). So does 
the first Lindbladian term in the second line of Eq.~(\ref{EQ:Lindblad_Fourier}).
Therefore the zero-momentum component experiences 
a non-trivial competition of Hamiltonian and dissipative dynamics.
The finite-$k$ components do nothing but decay. In particular, 
there is \emph{nothing in the Hamiltonian or Lindbladian
that can change the momentum of a given state}.
Therefore, in the long-time limit $t\rightarrow \infty$, it is legal to 
focus on the zero-momentum component, that is, the mean-field 
non-equilibrium steady states are actually the exact solution.

\subsection{Symmetries}
\label{eq:sym}
Eq.~(\ref{eq:master}) is invariant under a combined, discrete $\mathbbm{Z}_2$
symmetry transformation
\begin{align}
\mathbbm{Z}_2: [a+a^{\dagger},\sigma^x_{\ell},\sigma^y_{\ell}]\to [-(a+a^{\dagger}),(-\sigma^x_{\ell},-\sigma^y_{\ell})],
\label{eq:Z2}
\end{align}

which corresponds to a unitary transformation
\begin{align}
U_{\pi}=\exp (i\pi\left( a^{\dagger}a+\sum_{\ell=1}^N\sigma^z_\ell\right)).
\label{eq:Z22}
\end{align}
This symmetry is 
spontaneously broken at the Dicke superradiance transition.

Additionally, the spin sector of the Hamiltonian in Eq.\,\eqref{eq:hamiltonian} is invariant under a combination of time reversal $\mathcal{T}_{\ell}=-i\sigma^y_{\ell}K_\ell, \ t\rightarrow-t$ (for a spin $s=1/2$) and rotation in spin-space around the $y$-axis with angle $\theta=\pi$ denoted as $D^{1/2,\ell}_{y,\pi}=-i\sigma^y_{\ell}$, where $K_\ell$ is the complex conjugation operator such that $\mathcal{G}_{\ell}=D^{1/2,\ell}_{y,\pi} \mathcal{T}_{\ell}=-K_{\ell}$ and $\mathcal{G}_{\ell}\mathcal{G}^{-1}_{\ell}=1$ with $[\mathcal{G}_{\ell},\mathcal{G}^{-1}_m]=0$ and $[\mathcal{G}_{\ell},\mathcal{G}_m]=0$. If we write $G=\Pi_{\ell=1}^N\mathcal{G}_{\ell}$ we have
\begin{align}
GHG^{-1}=H.
\end{align}
In the absence of a loss channel in the spin sector, this means that the steady-state must be invariant under this transformation as well, which enforces $\langle\sigma^y\rangle=\langle G\sigma^yG^{-1}\rangle=-\langle\sigma^y\rangle\overset{!}{=}0$. This symmetry is broken in the presence of Liouvillian $\mathcal{L}_{\gamma}$ in the spin sector and therefore steady states with non-zero $\langle\sigma^y\rangle\neq0$ are accessible in the dynamics. In the photon sector, the corresponding symmetry is broken as well due to the presence of $\mathcal{L}_{\kappa}$, which leads to complex expectation values $\langle a\rangle\in\mathbb{C}$.

We mention here that the Hamiltonian dynamics together with the Lindblad contribution $\mathcal{L}_{\kappa}$ 
conserves the pseudo-angular momentum $\braket{\b{S}_t}^2$ but this conservation is explicitly 
broken by $\mathcal{L}_{\gamma}$.
Using semi-classical steady states defined below, one finds
%
\begin{align}
\partial_t \braket{\b{S}_t}^2&=2\braket{\b{S}_t}\cdot\partial_t\braket{\b{S}_t}\nonumber\\
&=-\gamma\left(\braket{\b{S}_t}^2+2\braket{\sigma^z_t}\l 1+\frac{\braket{\sigma^z_t}}{2}\r\right),
\label{EQ:CollectiveStateDecay}
\end{align}
such that the steady-state value requires 
$\lim\limits_{t \to \infty}\braket{\b{S}_t}^2= - 2\braket{\sigma^z}\l 1+\frac{\braket{\sigma^z}}{2}\r$ to hold.

\subsection{Experimental context in cavity QED}
\label{sec:ExperimentalContext}
\label{subsec:exp}
\begin{figure}
\includegraphics[width=9cm]{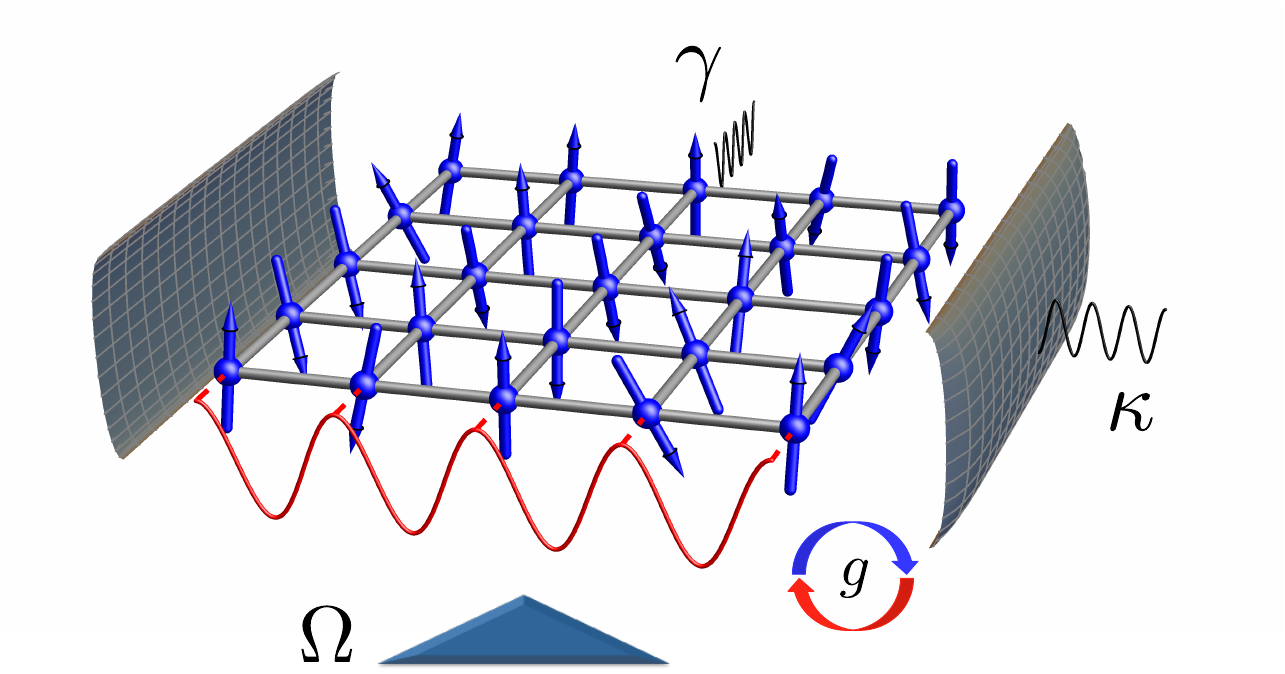}
\caption{Sketch of the system setup. The optical lattice inside the cavity is commensurate with the single light mode of the cavity that couples all atoms with strength $g$ to the light-field. The atoms can directly spontaneously decay with rate $\gamma$ into the reservoir of electromagnetic modes via the solid angle not covered by the cavity mirrors. These losses are uncorrelated single-site losses and collective emissions into a shared external reservoir once the system is in the superradiant state. The collective contribution to the decay is estimated in appendix \ref{App:SecCollectiveAtomicDecay} and found to be negligibly small for the current setup.  Photons can leave the system through the cavity mirrors with rate $\kappa$. The system is driven from the side with a laser of Rabi frequency $\Omega$ to stabilize the excitation number.}
\label{Fig:ExperimentalSetUp}
\end{figure}

To realize Eq.~(\ref{eq:hamiltonian}) in an 
optical cavity system, it is advantageous to suppress the motion of the atoms sufficiently such 
that the internal spin state dynamics dominates. To that end, an additional optical lattice 
potential inside the resonator has been realized recently by three different groups 
\cite{klinder15,landig16,neuzner16} paving the way for many-body quantum optics 
in which the relative phases of the emitters play a role. For the simple Dicke model 
the lattice and cavity mode function are engineered such that every atom 
couples with the same strength to the cavity photon, see Fig,\,\ref{Fig:ExperimentalSetUp}. Other arrangements, 
including mutually incommensurate periods \cite{habibian13}, 
can now be turned into experimental reality.

Moreover, some form of Raman-transition assisted pumping scheme is required 
to reach the strong-coupling regime for the effective spin-photon coupling $g$ 
needed to achieve the Dicke transition \cite{dimer07,baden14}.
There, the atomic levels to realize an effective spin system $\{\ket{\uparrow},\ket{\downarrow}\}$ can be the hyperfine-structure manifold of the ground states 
of $^{87}$Rb. Typically this is the $5^2$S$_{1/2}$ manifold.
The cavity-assisted Raman transitions are achieved by coupling to the states of the first excited state manifold $5^2$P$_{1/2}$ or $5^2$P$_{3/2}$. 

Moreover, the choice of laser frequencies in the experiment by Baden {\it et al.} \cite{baden14}
leads to the $\frac{U}{2N} \left(\sum_{\ell = 1}^N \sigma^z_\ell \right)a^\dagger a$ term in 
Eq.\,(\ref{eq:hamiltonian}) and we will come back to this experiment below 
in Subsec.\,\ref{subsec:singapore}. 

\subsection{Heisenberg-Langevin and mean-field master equations}
\label{subsec:langevin}
The Heisenberg equation of motion for an arbitrary system operator $A$ from the set $(\sigma^+_i,\sigma^-_i,\sigma^z_i)$, is calculated according to
\begin{align}
\partial_t A =& -i[H,A]+\gamma(1-\alpha)\sum_{\ell=1}^N \left( \sigma^{+}_{\ell}A\sigma^{-}_{\ell}-\frac{1}{2}\{\sigma^{+}_{\ell}\sigma^{-}_{\ell},A\}\right)\nonumber \\
&+\gamma \alpha \left( S^{+}A S^{-}-\frac{1}{2}\{S^{+}S^{-},A\}\right)
\label{EQ:HeisenbergDynamicswithLindblad},
\end{align}
where the Hamiltonian is given by Eq.\,\eqref{eq:hamiltonian} and $(\kappa,\gamma)$ refer to the cavity damping and the rate of spontaneous emission, respectively. For the atomic degrees of freedom we use the notation $\sigma^+_i=\ket{e}_i\bra{g},\sigma^z_i=\ket{e}_i\bra{e}-\ket{g}_i\bra{g}$ and $S^{\pm}=\sum_{\ell=1}^N \sigma^{\pm}_{\ell}$. Here $(e,g)$ refers to the excited and ground state of a two-level atom, respectively and $a$ labels the annihilation operator for a cavity photon. The Heisenberg-Langevin equations for these variables are:

\begin{align}
\partial_t a_t&=-\bigg[\kappa+i\l\omega_0+\frac{U}{2N}\sum_{\ell=1}^N\sigma^z_{\ell,t}\r\bigg]a_t\nonumber\\
&-i \frac{g}{\sqrt{N}} \sum_{\ell=1}^N\l\sigma^-_{\ell,t}+\sigma^+_{\ell,t}\r+\sqrt{2\kappa}a_{\rm in,t} \label{EQ:HeisenbergLangevinPhoton}\\
\partial_t \sigma^+_{i,t}&=i\left( \Delta+\frac{U}{N} a^{\dagger}_t a_t\right)\sigma^+_{i,t} -i\frac{g}{\sqrt{N}}\sigma^z_{i,t}\left(a_t+a^\dagger_t\right)\nonumber\\
& -\frac{\gamma}{2}\sigma^{+}_{i,t}+\frac{\gamma \alpha}{2}\sum_{\ell\neq i} \sigma^{+}_{\ell,t}\sigma^{z}_{i,t}+\mathcal{F}^+_{i,t}\\
\partial_t \sigma^z_{i,t}&=2\frac{g}{\sqrt{N}}\left(a_t+a^{\dagger}_t\right)\left(\sigma^-_{i,t} -\sigma^+_{i,t}\right)i\nonumber \\
&-(1+\sigma^z_{i,t})\gamma-\gamma \alpha \sum_{\ell\neq i}\left(\sigma^{+}_{i,t}\sigma^{-}_{\ell,t}+cc.\right)+\mathcal{F}^z_{i,t}
\label{EQ:HeisenbergLangevinPhotonSigmaz}
\end{align}
Here, $\left(a_{\rm in,t},\mathcal{F}^z_{i,t},\mathcal{F}^+_{i,t}\right)$ are the usual 
fluctuating quantum mechanical noise operators with zero mean. They result from integrating out the  bath of 
electromagnetic modes outside the cavity in the Born-Markov approximations, see App.~\ref{App:MakrovBath}.

If we take the bath to be in the vacuum state at zero-temperature, the noise-correlations for the atomic degrees of freedom can be expressed in the basis $(i,j)\in (+,-,z)$ by
\begin{align}
\braket{\mathcal{F}^{i}_{\ell',t'}\mathcal{F}^{j}_{\ell,t}}&=\gamma\delta(t-t')\bigg[\delta_{\ell,\ell'}+(1-\delta_{\ell,\ell'})\alpha\bigg]M^{ij}_{\ell'\ell}\\
M^{ij}_{\ell' \ell}&=2\left(
\begin{array}{ccc}
 0 & 0 & 0 \\
 \frac{1}{2}\sigma^{z}_{\ell',t}\sigma^z_{\ell,t'} & 0 & -\sigma^{z}_{\ell',t}\sigma^{-}_{\ell,t'} \\
  -\sigma^{+}_{\ell',t'}\sigma^{z}_{\ell,t} & 0 & 2 \sigma^{+}_{\ell',t'}\sigma^{-}_{\ell,t} \\
\end{array}
\right)_{ij}
\label{EQ:NoiseOperatorsCorrelsAtom}
\end{align}
where the indices $(i,j)\in(+,-,z)$ refers to the atomic variables. The expectation value averages over the bath degrees of freedom. Consequently, entries of the correlation matrix are still operator-valued. For contributions $\ell=\ell'$ the local operator algebra can be used to simplify correlations, see Eqs.~(\ref{EQ:CorrelationFunction}-\ref{EQ:CorrelationFunctionz}).
Similarly the noise-operators coupling to the photons are delta correlated in time
\begin{align}
\braket{a_{\rm in,t}a^{\dagger}_{\rm in,t'}}&=\delta (t-t')\label{EQ:NoiseCorrelator2}\;.
\end{align}
We will proceed to analyze the Heisenberg-Langevin equations in a mean-field framework where we apply a site-decoupling for the many-atom states. 
The mean-field state cannot keep track of the behaviour of atoms on different sites. Therefore, taking the expectation values $\braket{A}={\rm Tr} \left(A \otimes_{n=1}^N\rho_n\right)$,where the density matrix is site decoupled as $\rho=\otimes_{i=1}^N \rho_i$, leads to the mean-field equation for $N$ two-level atoms
\begin{align}
\partial_t \braket{a_t}=&-\bigg[\kappa+i\l\omega_0+\frac{U}{2}\braket{\sigma^z_t}\r\bigg]\braket{a_t} \nonumber \\
&-ig (\braket{\sigma^{-}_t}+\braket{\sigma^{+}_t})\label{EQ:MeanFieldSR1}\;,\\
\partial_t \braket{\sigma^+_t}=&\left[i\left( \Delta+U \braket{a^{\dagger}_t}\braket{a}_t\right)-\frac{\gamma}{2} (1-\beta\braket{\sigma^z_t})\right]\braket{\sigma^{+}_t}\nonumber\\
&-ig \braket{\sigma^z_t}(\braket{a_t}+\braket{a^{\dagger}_t})\;,\label{EQ:MeanFieldSR2}\\ \partial_t \braket{\sigma^z_t}
=&-\gamma\left[(1+\braket{\sigma^z_t})+2 \beta \braket{\sigma^{-}_t}\braket{\sigma^{+}_t}\right]\nonumber \\
&+2ig (\braket{a_t}+\braket{a^{\dagger}_t})(\braket{\sigma^{-}_t}-\braket{\sigma^{+}_t})\;.\label{EQ:MeanFieldSR3}
\end{align}
Here we have set $\beta=\alpha (N-1)$ which fixes the strength of the collective decay as discussed in Sec.\,\ref{subsec:hamiltonian}. In the remainder of the paper, we provide an analysis of these equations on a mean-field level and calculate the corresponding non-equilibrium steady-states 
and connect our theoretical results to the recent quantum optical realization of the Dicke phase transition with cavity-assisted Raman transitions \cite{baden14} to address the observed discrepancy between critical pump strengths and earlier calculations.
As pointed out above, we restrict our analysis to the case of weak collective decay contributions, with $0\leq\beta\ll 1$ which is an appropriate approximation for the present large sample limit, where the extensions of the cavity and the atomic sample are much bigger than the optical wavelength of the cavity modes.
In the superradiant phase, the collective emission of the atoms is locked to the cavity wave vector. In the present setting, the atomic resonance deviates strongly from the frequency associated to the cavity, i.e. $\Delta \gg c k_0$. Correspondingly, this energy mismatch suppresses the strength of the collective emission $\alpha$ even further.  
A conservative approximation of the geometric contribution $\alpha$ is given in App. \ref{App:SecCollectiveAtomicDecay}. We reserve an analysis of strong collective decay contributions for future work (see also Refs.~\cite{Scully2015}). 
\section{Results}
\label{sec:results}
In this section, we first compute an analytic formula for the 
critical coupling for the onset 
of Dicke superradiance in the presence of both correlated and uncorrelated atomic spontaneous emission.
We then use this formula to determine an effective atomic loss rate
for the experiment in Ref.~\cite{baden14}.
As we argued earlier the formula is in fact the \emph{exact solution} of the 
problem, in the "t-N"-limit even when the atomic loss is uncorrelated between 
individual sites. We also compare this effective decay rate to other atomic loss channels such as the collective polariton lifetime. We close by searching for signatures of the additional loss channel 
$\gamma$ in the cavity output spectrum and by computing the effective temperature 
of the system at the superradiance transition.
\subsection{Critical coupling for onset of superradiance \texorpdfstring{$g_c({\kappa,\gamma})$}{gc(kappa,gamma)}}
\label{subsec:critical}
We first transform Eqs.~(\ref{EQ:HeisenbergLangevinPhoton}-\ref{EQ:HeisenbergLangevinPhotonSigmaz}) into frequency space by the following relation:
\begin{align}
\mathcal{O}_t=\frac{1}{2\pi}\int\limits_{-\infty}^{\infty}e^{-i\nu t}\mathcal{O}(\nu) d\nu, \quad \mathcal{O}^{\dagger}_t=\frac{1}{2\pi}\int\limits_{-\infty}^{\infty}e^{-i \nu t}\mathcal{O}^{\dagger}(-\nu) d\nu\;,
\end{align}
where the operator $\mathcal{O}_t$ is either of $\left(a_t,\sigma^+_{i,t},\sigma^z_{i,t},a_{\rm in,t},\mathcal{F}^{+}_{i,t},\mathcal{F}^z_{i,t}\right)$ and $\mathcal{O}^{\dagger}_t$ refers to either of $\left(a^{\dagger}_t,\sigma^{-}_{i,t},a^{\dagger}_{\rm in,t}\right)$. 
We then specifically make a distinction between the semi-classical steady states and the amplitude fluctuations around these values by linearising Eqs.~(\ref{EQ:HeisenbergLangevinPhoton}-\ref{EQ:HeisenbergLangevinPhotonSigmaz}). We define the fluctuation operators in frequency space by the relation 
\begin{align}
\sigma^{+}(\nu)&=2\pi\braket{\sigma^+}\delta(\nu)+\delta\sigma^+(\nu), \label{EQ:Noise-Expansion1}\\
\sigma^{z}(\nu)&=2\pi\braket{\sigma^z}\delta(\nu)+\delta\sigma^z(\nu),\\
a(\nu)&= 2\pi\sqrt{N}\braket{a}\delta(\nu)+\delta a(\nu)\label{EQ:Noise-Expansion2}\;.
\end{align}
Where the set of steady-states $\braket{\sigma^+}$ and $\braket{\sigma^z}$ and $\braket{a}$ are solutions to Eqs.\,(\ref{EQ:MeanFieldSR1}-\ref{EQ:MeanFieldSR3}) in the long-time limit. Here, $\delta \sigma^+(\nu),\delta \sigma^{z}(\nu)$ and $\delta a(\nu)$ describe fluctuations about the semi-classical steady-state and $\delta(\nu)$ denotes a delta function in frequency space. The equations for the amplitude fluctuations are generated by  inserting Eqs.~(\ref{EQ:Noise-Expansion1}-\ref{EQ:Noise-Expansion2}) into the Fourier transformed set of Eqs.~(\ref{EQ:HeisenbergLangevinPhoton}-\ref{EQ:HeisenbergLangevinPhotonSigmaz}). At long times, we may neglect second-order terms in the fluctuations by assuming that the steady-state values are large compared to the associated fluctuations in the thermodynamic limit $N\to \infty$.
The linearized equations can be cast in matrix form
\begin{align}
\b{\mathcal{F}}(\nu)=\delta(\nu)f(\b{\sigma})+\b{G}^{-1}_{R}(\nu)\cdot \b{\delta \sigma}(\nu), \label{EQ:LinearLangevinEquation2}
\end{align}
where the fluctuation (noise) operators are collected in the vectors
 $\delta\b{\sigma}(\nu)$ ($\b{\mathcal{F}}(\nu)$):
\begin{align}
\scalemath{0.95}{\delta\b{\sigma}^{T}(\nu)}&\scalemath{0.95}{=\l \delta a(\nu), \delta a^{\dagger}(-\nu), \delta \sigma^{+}(\nu), \delta \sigma^{-}(-\nu), \delta \sigma^z(\nu) \r, }\\
\scalemath{0.95}{\b{\mathcal{F}}^{T}(\nu)}&\scalemath{0.95}{=\l \sqrt{2\kappa} a_{\rm in}(\nu),\sqrt{2\kappa} a^{\dagger}_{\rm in}(-\nu), \mathcal{F}^+(\nu), \mathcal{F}^{-}(-\nu), \mathcal{F}^{z}(\nu) \r.}
\end{align}
The inverse response function (retarded Green's function) then reads as
\begin{widetext}
\begin{align}
\b{G}_{\rm \b{R}}^{-1}(\nu)=\left(\scalemath{0.75}{
\begin{array}{ccccc}
 \frac{1}{2}i U \braket{\sigma^z}+\kappa -i \nu +i \omega_0 & 0 & i g & i g & i\frac{U}{2}\braket{a} \\
 0 & -\frac{1}{2} i U \braket{\sigma^z}+\kappa -i \nu -i \omega_0 & -i g & -i g & -i\frac{U}{2}\braket{a^{\dagger}} \\
 i g \braket{\sigma^z}-i U \braket{a^{\dagger}} \braket{\sigma^{+}}& i g \braket{\sigma^z}-i U \braket{a}  \braket{\sigma^+} & -i U \braket{a} \braket{a^{\dagger}}+\frac{\gamma }{2}(1-\beta \braket{\sigma^z})-i \Delta -i \nu  & 0 & ig (\braket{a}+\braket{a}^{\dagger}) \\
 i U \braket{a^{\dagger}}  \braket{\sigma^-}-i g \braket{\sigma^z} & -i g \braket{\sigma^z}+i U \braket{a} \braket{\sigma^{-}} & 0  & i U \braket{a}  \braket{a^{\dagger}}+\frac{\gamma }{2}(1-\beta \braket{\sigma^z})+i \Delta -i \nu & -ig (\braket{a}+\braket{a}^{\dagger}) \\
 2 i g (\braket{\sigma^{+}}-\braket{\sigma^{-}}) & 2 i g (\braket{\sigma^{+}}-\braket{\sigma^{-}}) & 2i g (\braket{a}+\braket{a}^{\dagger})+2\beta \gamma \braket{\sigma^-} & -2i g (\braket{a}+\braket{a}^{\dagger})+2\beta \gamma \braket{\sigma^+}  & \gamma -i \nu  \\
\end{array}}
\right)\;.
\label{Eq:GRinv}
\end{align}
\end{widetext}
The notation indicates that the responses of the system $\b{\delta\sigma}(\nu)$ to the "driving force" $\b{\mathcal{F}}(\nu)$ is indeed described by the function $\b{G}_{R}(\nu)$. The steady state contribution is encoded in $f(\b{\sigma})$.
We approach the phase transition from the normal phase, for which $\langle a \rangle = \braket{\sigma^+}=0$ 
and assume the atoms to be fully polarized $\langle \sigma^z \rangle =-1$.
Evaluating the condition for super radiance, $\lim\limits_{\nu \to 0}{\rm det} [\b{G}_{\rm \b{R}}^{-1}(\nu)]=0$, which is appropriate as long as $\beta<\sqrt{1+\left(\frac{\Delta}{\gamma/2}\right)^2}$, we find the critical coupling
\begin{align}
g_c(\gamma,\kappa,U,\beta)&=\frac{\sqrt{\frac{\gamma ^2}{4}(1+\beta)^2+\Delta ^2} \sqrt{\kappa ^2+\left(\omega_0-\frac{U}{2}\right)^2}}{2 \sqrt{\Delta  \left(\omega_0-\frac{U}{2}\right)}}.\;
\label{EQ:CriticalDickeCoupling}
\end{align}
This formula recovers the known expression \cite{dimer07} in the limit $(U,\gamma)\rightarrow 0$ and the critical coupling known from \cite{bhaseen12} in the limit $\gamma \to 0$. It can be seen that the spontaneous emission $\gamma$ ``shifts'' the atomic energy scale $\Delta$ and the photon loss rate $\kappa$ shifts the cavity frequency $\omega_0$. As expected, the addition of spontaneous atomic emission leads
to an increased value for the critical coupling $g_c$. Comparing this value Eq.~(\ref{EQ:CriticalDickeCoupling}) 
for the critical coupling to Eq.\,(103) in Ref.~\onlinecite{dalla13}, 
we note the structural similarity. However, the prefactor is different and this is due to 
the different nature of atomic decay processes included. Here, we included 
the conventional directed spontaneous emission into the spin-down state, while 
Ref.~\onlinecite{dalla13} included a dissipative dephasing term.
\subsection{Comparison with Singapore experiment}
\label{subsec:singapore}
In the Baden {\it et al.} experiment, the Dicke model was realized 
using cavity-assisted Raman transitions \cite{baden14}. 
A sudden increase in the number of detected cavity photons upon 
ramping up the drive strength of an external laser has been
associated with the threshold for Dicke superradiance. 
Some elements of the experiment we have already 
mentioned above in Subsec.~\ref{subsec:exp}.

\begin{figure}
\includegraphics[width=8.5cm]{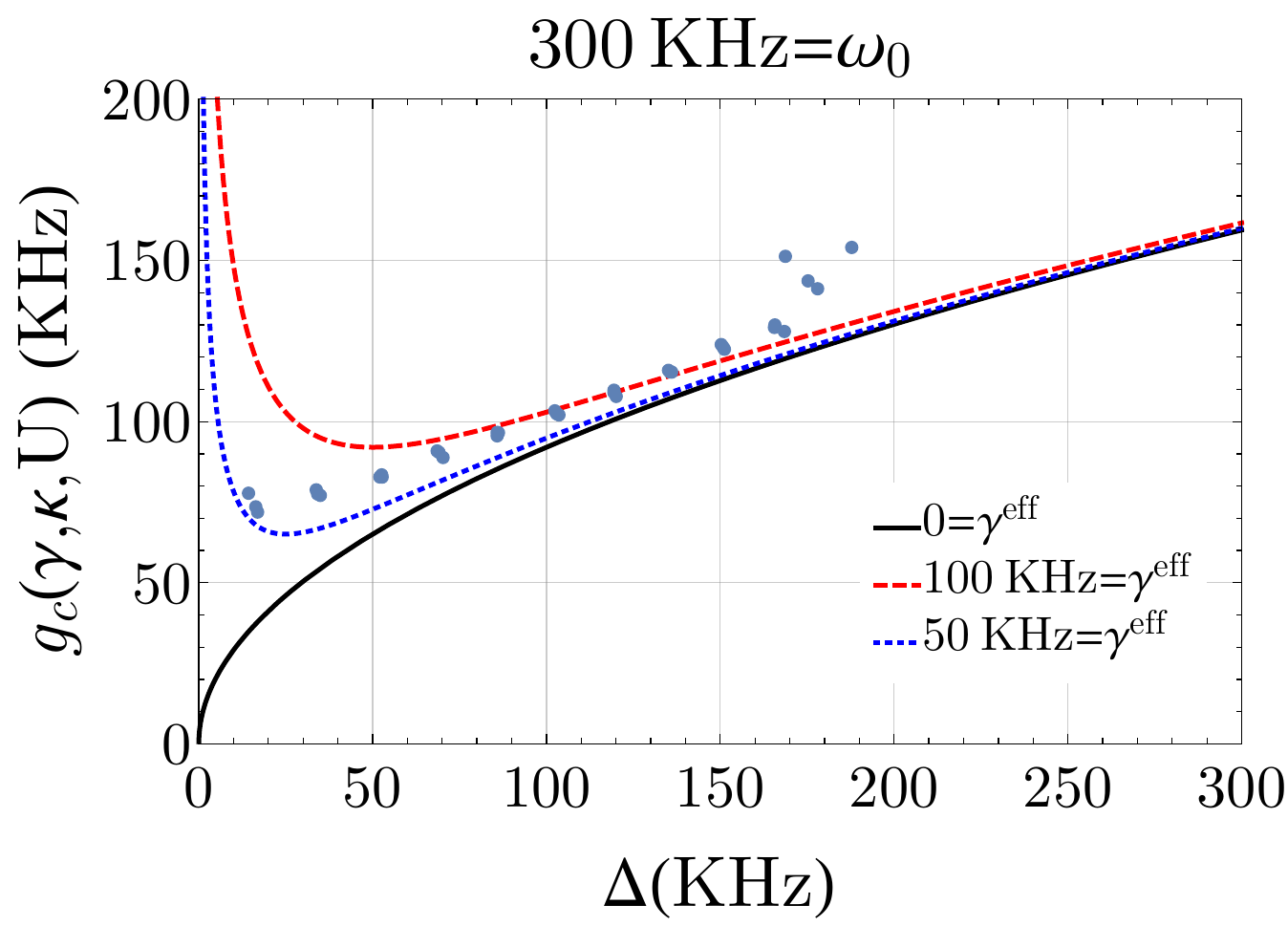}
\caption{
Critical coupling strength for the onset of superradiance. We compare Eq.\,\eqref{EQ:CriticalDickeCoupling} with $\beta \to 0$ to the most recent data (dots) 
from the Singapore experiment. We give an estimate for a lower (dotted) and upper (dashed) effective decay rate $\gamma^{\rm eff}$ (using the Lindblad of Eq.\,\ref{EQ:LindbladOperatorgamma}) of the effective atomic dipole and compare it to the theory curve for $\gamma^{\rm eff}=0$ (solid). $\gamma^{\rm eff}$ should be regarded as an effective spin decay rate after eliminating the far- detuned excited state, possibly other decay channels, and other experimental imperfections. Where we have used the experimentally determined values $U=-12.4$ KHz and $\kappa=100$ KHz. In the regime $|\omega_0|\gg |U|$ the effects of the frequency shift for the critical coupling are negligible. For small longitudinal fields $\Delta$ the theory curves show a clear upswing in the critical coupling, see text for details.
}
\label{fig:fit}
\end{figure}

Baden {\it et al.} \cite{baden14} compared the experimentally observed 
threshold couplings to the conventional theory value 
without spontaneous emission \cite{dimer07,bhaseen12} and found a discrepancy: 
higher pump strengths than predicted were necessary to observe 
an increase in photon numbers.
Using the computed value for the critical coupling in Eq.\,\eqref{EQ:CriticalDickeCoupling}, in the limit $\beta \to 0$ (for a conservative approximation see App. \ref{App:SecCollectiveAtomicDecay}), we may determine an effective atomic decay rate $\gamma=\gamma^{\rm eff}$, see Fig.\,\ref{fig:fit}, to reduce the discrepancy between experiment and theory. An additional interesting regime to pin down the effects of $\gamma^{\rm eff}$ 
is the critical region for small longitudinal spin detuning $\Delta$.
From the critical coupling Eq.~(\ref{EQ:CriticalDickeCoupling}), we 
observe that $g_c(\gamma^{\rm eff},\kappa,U)$ becomes large for small $\Delta$ 
provided $\gamma^{\rm eff}$ is finite. By contrast in the strict $\gamma^{\rm eff}\rightarrow 0$ 
limit $g_c(\gamma^{\rm eff} = 0, \kappa,U)$ {\textit{decreases}} for small $\Delta$. 
New rounds of data-taking can access this regime 
with improved accuracy \footnote{M. Barrett/Singapore Experiment, Private Communication}.

We find it useful to use a single effective atomic decay, captured by the Lindblad operator of Eq.\,\eqref{EQ:LindbladOperatorgamma}, to enable an experiment-theory comparison. However, other sources of noise (noise in the trapping potential, loss of atoms from the trap, dissipative dephasing, noise in the driving laser) could also be modeled and included.

To this end, one may wonder whether $\gamma^{\rm eff}$ can be explained by the decay rate that the  $\ket{\uparrow}=\ket{F=2,m_{F}=2}$ state in the $5^2$S$_{1/2}$ manifold inherits from the excited state in the $5^2$P$_{3/2}$ manifold to which the cavity couples. 
This inherited decay rate can be estimated as
\begin{align}
\gamma_{\rm inherited}&=\chi \l\Omega_r /\Delta_r\r^2 \gamma_{\rm exc}=\chi g^2/(NC \kappa)\;,
\label{EQ:decayeff}
\end{align}
where the proportionality constant $\chi$ is fixed by the transition strengths between the involved atomic levels.  $C=g^2_{\rm cav}/(\kappa\gamma_{\rm exc})$ is the single atom cooperativity and $(g_{\rm cav},\kappa,\gamma_{\rm exc})=2\pi \times (1.1,0.1,3)$ MHz. $\kappa$ is the cavity decay rate and $2g_{\rm cav}$ is the single photon Rabi frequency for the transition of $\ket{F=2,m_{F}=2}$ to $\ket{F'=3,m_F'=3}$, $\gamma_{\rm exc}$ is the fundamental atomic decay rate of the excited state level coupling to the cavity. $\Omega_r$ and $\Delta_r$ are the Rabi frequency and the detuning of the driving laser, respectively, that couples the $\ket{\uparrow}$ state to the aforementioned excited state of the $5^2$P$_{3/2}$ manifold. We have set the Raman coupling strength $g = \sqrt{N}g_{\rm cav}\frac{\Omega_r}{\Delta_r}$ in Eq.\,\eqref{EQ:decayeff} with $N$ a fixed number of atoms in the ground-state manifold.

\begin{table}[h!]\centering 
\begin{tabular}{c | c | c | c} 
 $\gamma_{\rm exc}$ & $\gamma^{\rm eff}(g=g_c)$ & $\gamma_{\rm Polariton}(g=g_c)$  & $\gamma_{\rm inherited}(g=g_c)$ \\ \hline
 3000 KHz & $(50-100)$KHz & $(72-103)$ KHz & 0.02 KHz
 \end{tabular} 
  \caption{Overview of various decay rates and their KHz values for the experimental setup in \cite{baden14}. The range for $\gamma^{\rm eff}(g=g_c)$ is estimated in Fig.\,\ref{fig:fit}. $\gamma_{\rm Polariton}$ is calculated from Eq.\,\eqref{EQ:GammaPolariton} with the set of parameters $(\omega_0,\kappa,\Delta,\gamma^{\rm eff})$=$(300,100,150,(50;100))$KHz. $\gamma_{\rm inherited}$ is determined by the model parameters and expressed as $\gamma_{\rm inherited}(g=g_c)=24 g^2_c/(NC \kappa)$, where $N$ is the number of atoms in the trap, $C=g^2_{\rm cav}/(\kappa,\gamma_{exc})$ is the single atom cooperativity, $\kappa$ is the cavity decay rate and $2g_{\rm cav}$ is the single photon Rabi frequency for the transition of $\ket{F=2,m_{F}=2}$ to $\ket{F'=3,m_F'=3}$ transition. The value for $\gamma_{\rm inherited}$ was calculated for $(N,C,g_c,\kappa)=(5\cdot 10^4,4,120 {\rm KHz},100 {\rm KHz})$ where the number of atoms represents a typical order of magnitude for the experiment in\cite{baden14}.}
  \label{tab:one}
 \end{table}
As we show and discuss in Table \ref{tab:one}, $\gamma_{\rm eff}\gg \gamma_{\rm inherited}$, such that $\gamma_{\rm inherited}$ alone is not sufficient to explain the experimental data. The conservative estimate for an upper bound for the geometrical factor $\beta\approx 10^{-3}$ shows that the collective atomic decay channel is irrelevant for the present setup, justifying the restriction of the following analysis to the case $\beta=0$.

\subsection{Polariton decay rates}
\label{subsec:collectivedecayrates}
For non-zero atom-light coupling, the atomic and photonic excitations of the system start to hybridize and are commonly referred to as polaritons. The decay of the polaritons describes a correlated decay mechanism with rate $\gamma_{\rm polariton}$, which involves many atoms and photons. The corresponding decay rate is a function of the bare decay rates of the individual atoms $\gamma$, the bare decay rate of the individual photons $\kappa$, the energies of the bare atoms $\Delta$ and of the photons $\omega_0$ as well as the atom photon coupling $g$. The rates can be read off from the imaginary part of the resonance frequencies $\nu$ for the linearized system dynamics, that can be determined from 
\begin{align}
\lim_{(\beta,U)\to 0}{\rm det}[\b{G}^{-1}_{\rm R}(\nu)]&=0\\
\gamma_{\rm Polariton}&=|{\rm Im}(\nu)|
\label{EQ:AtomicResonances}
\end{align} 

\begin{figure*}
\subfloat[]{\includegraphics[width=82mm]{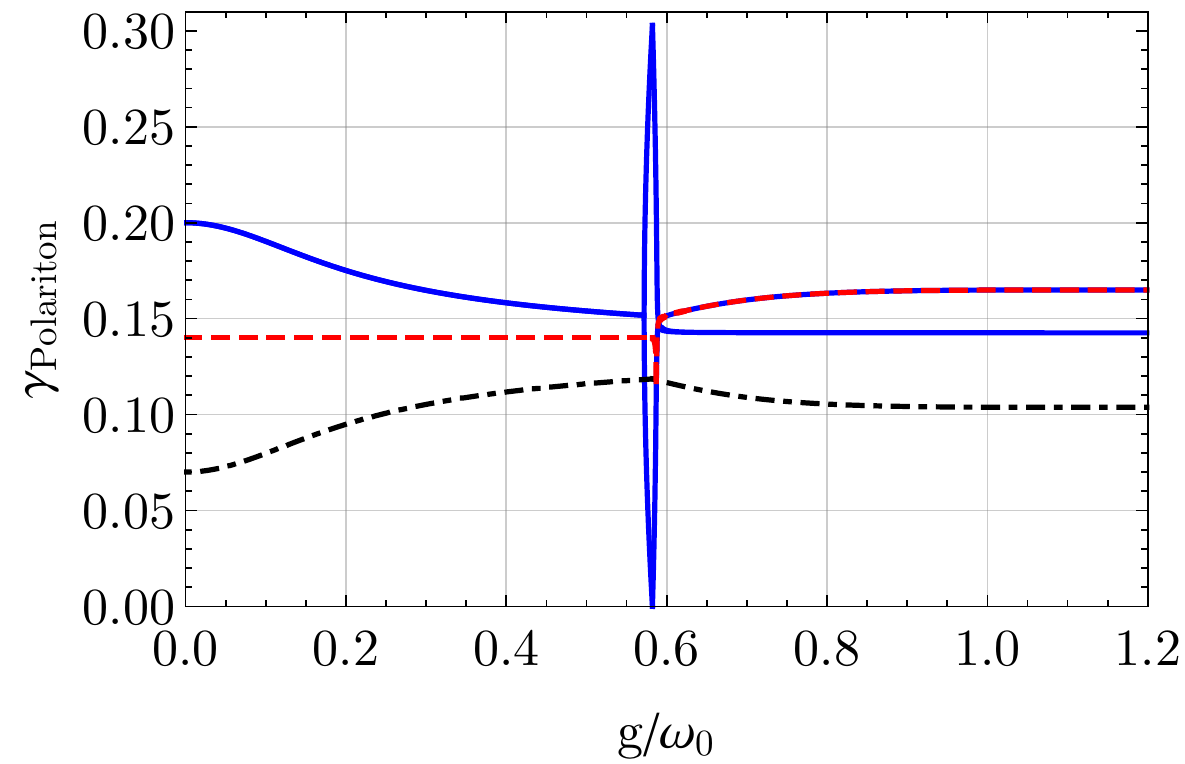}\label{subfig:CollectiveDecayBig}}
\subfloat[]{\includegraphics[width=80 mm]{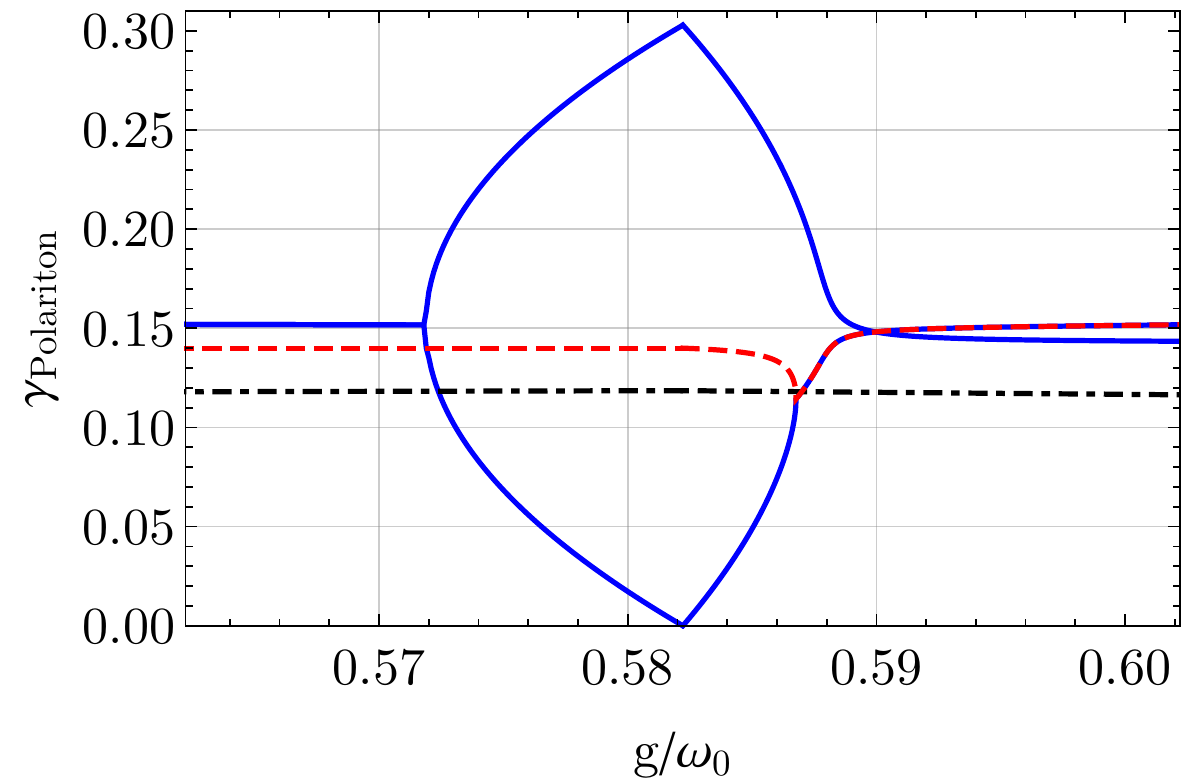}\label{subfig:CollectiveDecaySmall}}
\caption{Decay rates $(\gamma_{\rm Polariton})$ of (in general) hybridized atom-photon modes for varying atom-light coupling strength $g/\omega_0$ obtained from solving Eq.\,\eqref{EQ:AtomicResonances} for the set $\Delta/\omega_0=1.3, \kappa/\omega_0=0.2, \gamma/\omega_0=0.14$. (a) For $g \to 0$, the lifetimes are given by the microscopic cavity decay rate $\kappa/\omega_0=0.2$ (solid) and the single atom decay rate $\gamma/(2\omega_0)=0.07$ for the atomic polarisations (dot-dashed) and the decay rate associated to the density of excitations $\gamma/\omega_0=0.14$ (dashed).
(b) Close-up around the critical atom-light coupling strength $g=g_c\approx 0.582 \omega_0$. In the regime $g<g_c$ there is a splitting of a polariton branch into a mode with finite and vanishing lifetime at the phase transition point $g=g_c$. The splitting occurs when the real-part of the excitation frequencies $\nu$ vanishes (not shown).}
\label{Fig:PolaritonDecayRates}
\end{figure*}

In Fig.\,\ref{Fig:PolaritonDecayRates} we plot the effective decay rates $\gamma_{\rm polariton}$ as a function of the atom-light coupling $g$ as they are also shown in \cite{dimer07} for the $\kappa$ only case. 
As such $\gamma_{\rm Polariton}$ sets the width of the resonance peaks in the cavity-spectrum. We explore certain limits for the collective decay rates.
For $g \to 0$,  the resonances calculated from Eq.\,\eqref{EQ:AtomicResonances} are located at $\nu_{\rm atom}=\pm\Delta-i \frac{\gamma}{2}$, at $\nu_{\rm photon}=\pm\omega_0-i\kappa$ and the $\sigma_z$-resonance is at $\nu=-i\gamma$, see Fig.\,\ref{subfig:CollectiveDecayBig}. Corresponding to resonances located at the characteristic atom and photon frequencies with a line-shape of a Lorentz-curve with a width determined from the microscopic decay rates.
At $g=g_c$ the decay rate of the critical pole with finite imaginary part (see Fig.\,\ref{subfig:CollectiveDecaySmall}, solid, blue line) is given as
\begin{align}
\gamma_{\rm Polariton}(g=g_c)&= \frac{2  \left(\gamma ^2 \kappa +2 \gamma  \left(\kappa ^2+\omega_0^2\right)+4 \Delta ^2 \kappa \right)}{\gamma ^2+8 \gamma  \kappa +4 \left(\Delta ^2+\kappa ^2+\omega_0^2\right)} \label{EQ:GammaPolariton}\\
\lim_{\gamma \to 0}\gamma_{\rm Polariton}(g=g_c)&=\kappa\frac{2 \Delta ^2 }{\Delta ^2+\kappa ^2+\omega_0^2},
\end{align}

\label{subsec:conn}
%

\subsection{Non-equilibrium steady states for spins and photons}
\label{subsec:steady}

In this section, we discuss the steady-state operator expectation values $\braket{a},\braket{\sigma^+},\braket{\sigma^z}$,
where $\braket{a}$ is the complex field amplitude that accounts for a coherent photon condensate, $\braket{\sigma^+}$ is the complex atomic polarization amplitude and $\braket{\sigma^z}$ measures the atomic population imbalance. 
The dynamics of the expectation values is given by the mean-field equations Eqs.~(\ref{EQ:MeanFieldSR1}-\ref{EQ:MeanFieldSR3}). In the semi-classical picture for a spin-$1/2$ system we construct the expectation value of the spin-vector $\braket{\b{S}_t}=(\braket{\sigma^x(t)},\braket{\sigma^y(t)},\braket{\sigma^z_t})^T$. It defines the orientation of the averaged atomic Bloch vector.
The non-equilibrium Bloch dynamics of the collective angular momentum without 
spontaneous emission was studied in Ref.~\onlinecite{bhaseen12}.

\begin{figure*}
\subfloat[]{\includegraphics[width=80mm]{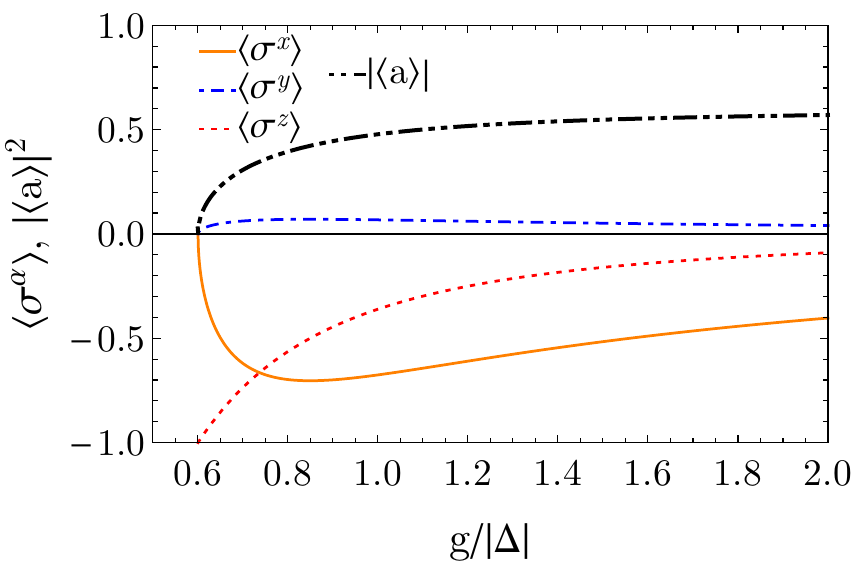}\label{subfig:Steady-State-Values}}
\subfloat[]{\includegraphics[width=78mm]{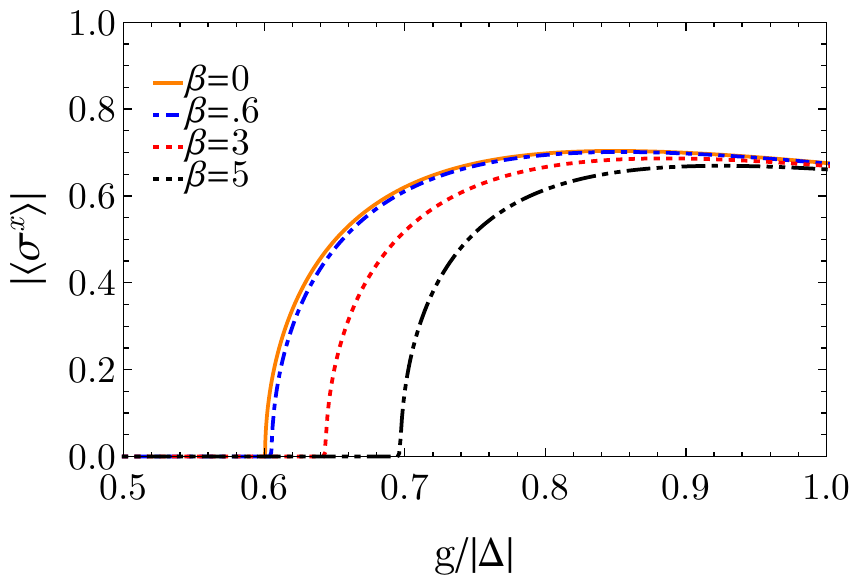}\label{subfig:collectivedecaysmall}}
\caption{Stable steady-state field amplitudes in the superradiant phase for the set of parameters $\gamma=\kappa=0.2 |\Delta|, \omega_0=1.4 |\Delta|$. The critical coupling $g_c$ is given by Eq.\,\eqref{EQ:CriticalDickeCoupling} which is valid as long as $\beta <\sqrt{1+\left(\frac{\Delta}{\gamma/2}\right)^2}\approx 10$. (a) Amplitudes $|\braket{a}|$ and $\left(\braket{\sigma^x},\braket{\sigma^y},\braket{\sigma^z}\right)$ without a collective decay contribution, i.e.\,$\beta=0$. The critical coupling evaluates to $g_c(\beta=0)/\Delta\approx 0.6$.
(b) Influence of collective decay processes of strength $\beta=\alpha(N-1)$ for steady-state field amplitude $|\braket{\sigma^x}|$.  }
\label{Fig:SteadyStateswithandwithoutcollectivedecay}
\end{figure*}

An analytical solution for the semi classical steady-states $(\partial_t \braket{\sigma^{\alpha}_t})=0$ and $\partial_t \braket{a_t}=0$ is accessible by setting $U=0$ and $\beta=0$. However, we show the effect of the collective decay contribution for the steady-state of the $\braket{\sigma^x}$ order parameter in Fig.\,\ref{subfig:collectivedecaysmall}.
Analysis of effective Dicke Hamiltonians with non-zero $U$ in the $\gamma \to 0$ regime have been investigated in detail \cite{Keeling10,bhaseen12,Nagy2008} such that we focus on consequences of non-vanishing radiative decay. We mention that in an experimental realization with cavity-assisted Raman transitions $U=0$ is achieved by having equal amplitudes for co- and counter-rotating terms of the effective Dicke-Hamiltonian \cite{dimer07}. Slight experimental mismatches of the amplitudes lead to vanishingly small $|U|\ll (|\omega_0|,|\Delta|,\kappa,\gamma)$ such that the resulting nonlinearities in the equation of motion can safely be neglected.  
We solve the system of non-linear equations for the fixed points to obtain the steady-states. 
For $g<g_c$ the only steady-state is $\braket{a}=\braket{\sigma^+}=0$ and $\braket{\sigma^z}=-1$. This is the empty atom-cavity system as the spontaneous atomic decay and photon loss depletes the system of all excitations. 
The mean-field expectation values for the fields in the superradiant phase $g>g_c$ are:
\begin{align}
\braket{a}&=\pm\frac{\sqrt{\frac{\kappa ^2+\omega_0^2}{\omega_0}} \sqrt{\Delta  \left(1-\frac{J_c}{J}\right)}}{\sqrt{2} (-\omega_0+i \kappa )},\label{EQ:SSa}\\
\braket{\sigma^x}&=\left(\braket{\sigma^+}+\braket{\sigma^-}\right)=\pm\frac{\sqrt{\Delta  (J-J_c)}}{\sqrt{2} J},\\
\braket{\sigma^y}&=-i\left(\braket{\sigma^+}-\braket{\sigma^-}\right)=\mp\gamma \frac{\sqrt{\Delta  (J-J_c)}}{2 \sqrt{2} \Delta  J}\label{EQ:Ssy},\\
\braket{\sigma^z}&=-(J_c/J).\label{EQ:SSz}
\end{align}
Here the different signs for the steady-state solutions reflect the $\mathbbm{Z}_2$ symmetry which is spontaneously broken by the choice of a specific state, see Eq.\,\eqref{eq:Z2}, and we have abbreviated the notation by defining
\begin{align}
J&=\frac{g^2 \omega_0}{\kappa^2+\omega_0^2},\label{EQ:J}\\
J_c&=\frac{\gamma ^2+4 \Delta ^2}{16 \Delta }.\label{EQ:Jc}
\end{align}
A plot of Eqs.~(\ref{EQ:SSa}-\ref{EQ:SSz}) is given in Fig.~\ref{subfig:Steady-State-Values}.
The critical coupling strength $g_c$ for the superradiant phase transition (see Eq.\,\eqref{EQ:CriticalDickeCoupling} in the $(\beta,U)\to 0$ limit) can also be obtained by equating Eqs.~\eqref{EQ:J} and \eqref{EQ:Jc}.
For $\Delta<0$ there is no real-valued solution for 
the magnetizations $(\braket{\sigma^x},\braket{\sigma^y})$ which means that this regime excludes a stable photon condensate.

Note that the solutions for the mean-field expectation values do not recover the solutions that are obtained by taking the  $\gamma \to 0$ limit from the outset in Eqs.~(\ref{EQ:MeanFieldSR1}-\ref{EQ:MeanFieldSR3}). This is because the present steady-state is usually approached with a rate $\propto 1/\gamma$ which diverges in the $\gamma \to 0$ limit.


\subsection{Cavity output spectrum}
\label{sec:CavityOutPut}
The internal dynamics of the atom-cavity system can be probed by analyzing the light that leaks from the cavity mirrors. 
We employ standard input-output theory for the quantum Langevin equations \cite{collett84,gardiner85,dimer07} to calculate the cavity spectrum in the limit $(U,\beta)\to 0$, to focus on the effect of single-site atomic spontaneous emission. 
The input-fields are related to the output fields by the relation
\begin{align}
a_{\rm out}(\nu)&=\sqrt{2 \kappa}a(\nu)-a_{\rm in}(\nu),\label{EQ:InputOutputRelation}\\
a_{\rm out}^{\dagger}(-\nu)&=\sqrt{2 \kappa}a^{\dagger}(-\nu)-a_{\rm in}^{\dagger}(-\nu).
\end{align}
The annihilation operators $\left(a_{\rm out}(\nu),a_{\rm in}(\nu),a(\nu)\right)$ correspond to the output field, the input field, and the intra cavity field, respectively. For a vacuum field input, the correlations of the noise operators as given by the set of  Eqs.~(\ref{EQ:NoiseOperatorsCorrelsAtom}-\ref{EQ:NoiseCorrelator2}) in frequency space are
\begin{align}
\braket{a_{\rm in}(\nu')a^{\dagger}_{\rm in}(-\nu)}&=\delta(\nu+\nu')\;,\label{EQ:CorrelationFunction}\\
\braket{\mathcal{F}^-(-\nu')\mathcal{F}^+(\nu)}&=\gamma \delta(\nu+\nu')\;,\\
\braket{\mathcal{F}^{z}(\nu)\mathcal{F}^z(\nu')}&=2\gamma(1+\braket{\sigma^z})\delta(\nu+\nu')\;,\\
\braket{\mathcal{F}^{z}(\nu)\mathcal{F}^{+}(\nu')}&=2\gamma\braket{\sigma^+}\delta(\nu+\nu').\label{EQ:CorrelationFunctionz}
\end{align}
We solve Eq.\,\eqref{EQ:LinearLangevinEquation2} for the fluctuations around the photon condensate and omit the coherent contribution coming from the zero-frequency components specified by $f(\b{\sigma})$.
Making use of Eqs.~(\ref{EQ:InputOutputRelation}-\ref{EQ:CorrelationFunction}), the cavity fluorescence spectrum $S(\nu)$ (for a vacuum input field) accounting for the fluctuations around the steady state is
\begin{align}
S(\nu)&=\braket{a^{\dagger}_{\rm out}(\nu)a_{\rm out}(\nu)}
=2\kappa\braket{\delta a^{\dagger}(\nu)\delta a(\nu)}\nonumber \\
&=2\kappa\int_{-\infty}^{\infty}e^{-i\nu \tau}\braket{\delta a^{\dagger}(0)\delta a(\tau)}d\tau.
\label{Eq:CavSpectrumRelation}
\end{align}
\begin{figure*}
\subfloat[]{\includegraphics[width=8cm]{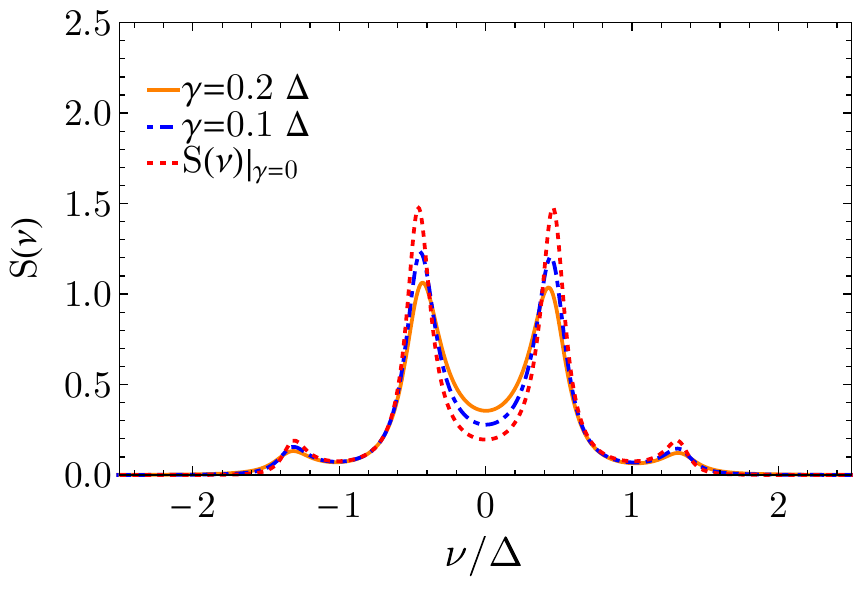}}
\subfloat[]{\includegraphics[width=8cm]{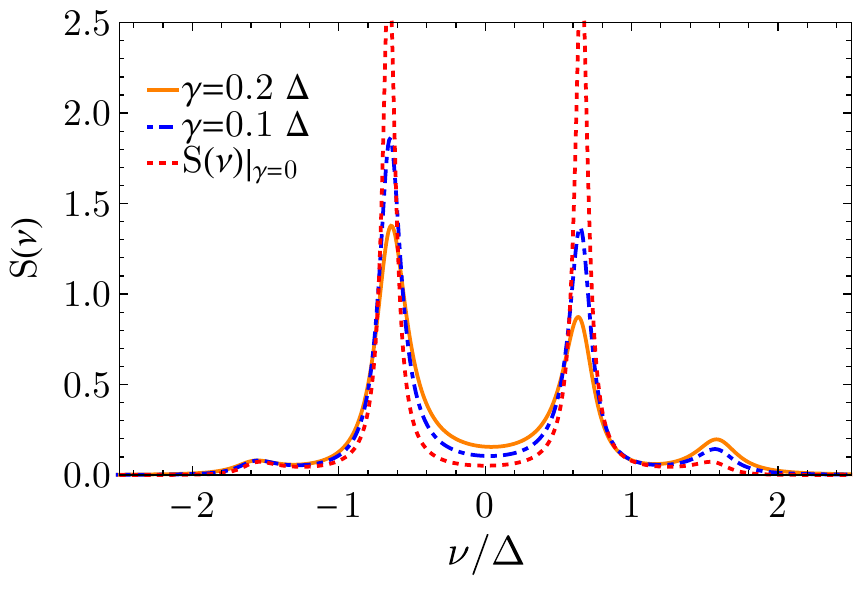}}
\caption{
Cavity spectra in the vacuum phase with spontaneous emission and a 
broken frequency symmetry $S(\nu)\neq S(-\nu)$. We have normalized the spectra such that $\int S(\nu)d\nu=1$. When there is no atomic spontaneous emission $(\left.S(\nu)\right|_{\gamma=0})$, the frequency symmetry of the cavity spectrum is restored.
All parameters are in units of $|\Delta|$: $\kappa=0.2 |\Delta|, g=0.4 |\Delta|<g_c$. The figure (a) shows the on-resonance spectrum: $\omega_0=1.0|\Delta|$, the figure (b) shows the off-resonance spectrum at $ \omega_0=1.4|\Delta|$.}
\label{fig:output_vacuum}
\end{figure*}

\begin{figure*}
\subfloat[]{\includegraphics[width=8cm]{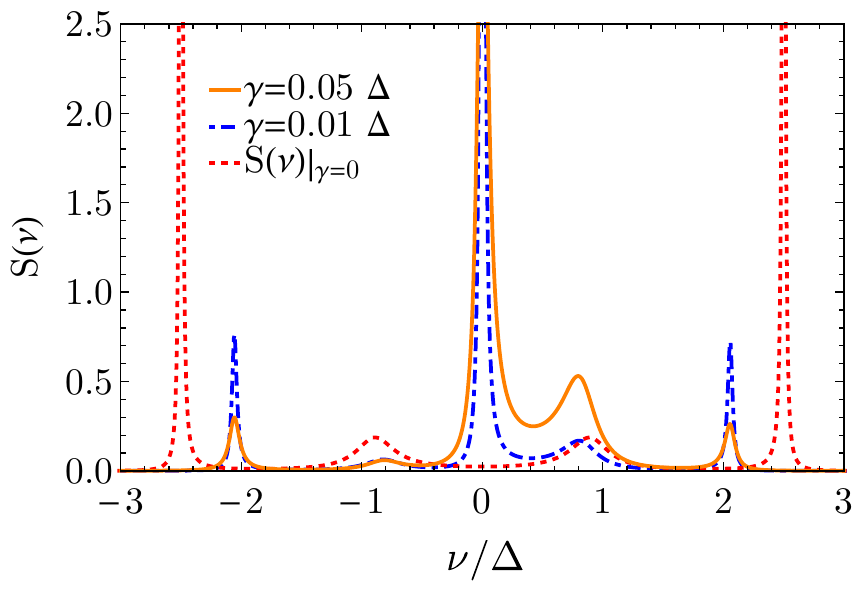}}
\subfloat[]{\includegraphics[width=8cm]{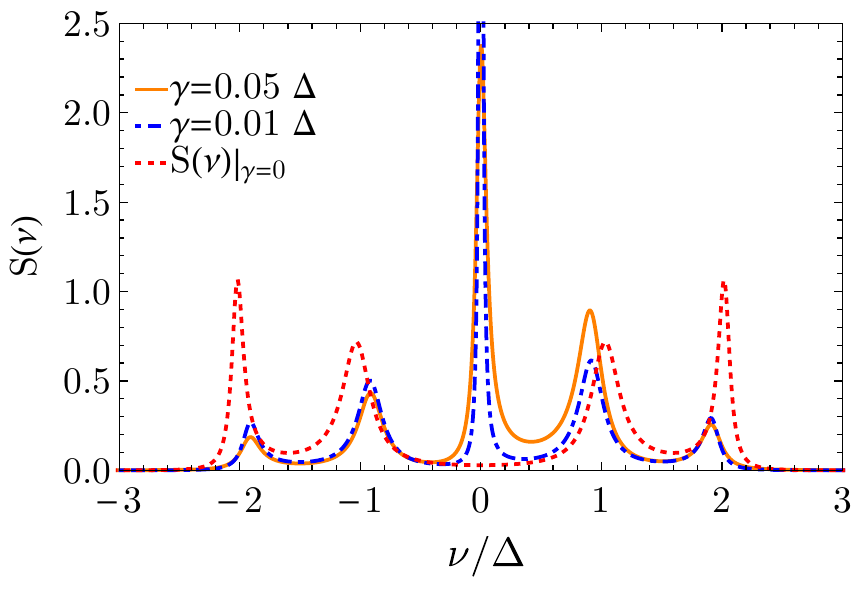}}
\caption{
Cavity spectra in the superradiant regime with spontaneous emission and a 
broken frequency symmetry $S(\nu)\neq S(-\nu)$. We have normalized the spectra such that $\int S(\nu)d\nu=1$.
All parameters are in units of $|\Delta|$: $\kappa=0.2 |\Delta|, g=0.8 |\Delta|>g_c$. Here, $\left.S(\nu)\right|_{\gamma=0}$ is understood as solving the Eqs.~\eqref{EQ:HeisenbergLangevinPhoton}-\eqref{EQ:HeisenbergLangevinPhotonSigmaz} on a mean-field level by setting $\gamma=0$ from the outset and calculating $S(\nu)$ with Eq.\,\eqref{Eq:CavSpectrumRelation}, see \cite{dimer07}. In figure (a) the system is on resonance: $ \omega_0=|\Delta|$, in figure (b) it is off resonance: $ \omega_0=1.4|\Delta|$.}
\label{fig:output_superradiant}
\end{figure*}

The cavity spectrum in the $g<g_c$ case for the steady-states ${\braket{\sigma^+}=\braket{\sigma^-}=0}$ and ${\braket{\sigma^z}=-1}$
becomes
\begin{align}
S(\nu)=&8 g^2 \kappa \frac{ s(\nu)}{|\Omega(\nu)|^2},\label{EQ:CavSpectrumDarkState}\\
s(\nu)=&\left(\gamma  \left(\gamma ^2+4 (\Delta -\nu )^2\right) \left(\kappa ^2+(\nu +\omega_0\right)^2+32 \Delta ^2 g^2 \kappa \right)\\
\Omega(\nu)=&(\kappa -i \nu )^2 \left(4 \Delta ^2+(\gamma -2 i \nu )^2\right)\nonumber\\
&+\omega_0^2 
\left(4 \Delta ^2+(\gamma -2 i \nu )^2\right)-16 \Delta  g^2 \omega_0.
\end{align}
The poles in Eq.\,\eqref{EQ:CavSpectrumDarkState} correspond to the hybridized atom-cavity eigenmodes of the system. They are given by the solutions to the equation ${\rm det} [\b{G}_{\rm \b{R}}^{-1}(\nu)]=0$ where $\b{G}_{\rm \b{R}}^{-1}(\nu)$ is defined in Eq.\,\eqref{Eq:GRinv}.
Close to the superradiance transition, two poles become purely imaginary and characterize the overdamped dynamics at the phase transition, see e.g.\,\citep{dalla13}. At the transition a single critical mode approaches the origin linearly in $(g-g_c)$. The corresponding equations for $\gamma\neq0$ can be found in the appendix, Eqs.~(\ref{EQ:CriticalPole1}-\ref{EQ:CriticalPole2}).

The output spectrum in the presence of spontaneous atomic decay is no longer symmetric under inversion on the frequency axis $S(\nu)\neq S(-\nu)$, see Fig.\,\ref{fig:output_vacuum}. This is due to the fact that in the presence of atomic decay, cavity photons can exit the cavity in two different ways. Either directly via the cavity decay channel $\sim \kappa$ or indirectly via exciting an atom and subsequently decaying via spontaneous emission $\sim \gamma$. The latter process of combined excitation and decay prefers photon states with positive frequency. This leads to a reduction of $S(\nu)$ for positive frequencies and introduces the mentioned asymmetry in the photon output spectrum.
In the limit of vanishing spontaneous emission, the cavity spectrum collapses to the familiar result 
\cite{dimer07} and the frequency symmetry is restored:
\begin{align}
\lim_{\gamma \to 0}S(\nu)&=\frac{16 \Delta ^2 g^4 \kappa ^2}{|\Omega'(\nu)|^2},\\
\Omega'(\nu)&=(\Delta^2-\nu^2) (\kappa -i \nu )^2+\omega_0^2 (\Delta^2-\nu^2)-4 \Delta  g^2 \omega_0.
\end{align}

A typical cavity output spectrum for the superradiant case $g>g_c$ is defined in Eq.\,\eqref{Eq:CavSpectrumSR} and can be seen in Fig.\,\ref{fig:output_superradiant}. Here $\gamma>0$ leads to a broadening of the spectrum and a pronounced weight of $S(\nu)$ at positive frequencies $\nu>0$ due to the dominant effect of stimulated emission and absorption over spontaneous decay effects.
 In Fig.\,\ref{fig:output_superradiant} and Fig.\,\ref{fig:output_vacuum} the expression $S(\nu)|_{\gamma=0}$ refers to the cavity spectrum that is obtained by setting $\gamma=0$ in Eqs.~\eqref{EQ:HeisenbergLangevinPhoton}-\eqref{EQ:HeisenbergLangevinPhotonSigmaz} and by  then following the same procedure as outlined above, see \cite{dimer07}.

%

\subsection{Effective temperature}
\label{subsec:Teff}
In Ref.~\onlinecite{dalla13}, the authors outlined an approach to extract 
the effective temperature of open quantum-optical systems. The idea 
is to map the photon equation of motion to classical Langevin equations 
and read off the effective temperatures as a function of the noise correlation functions.
Here, we generalize this analysis to include single-site atomic spontaneous emission $\gamma$ and extract the corresponding effective temperature.
To that end, we define the real part of the photon component $\delta x(\nu)$ and 
the corresponding noise operator $\mathcal{F}_x(\nu)$:
\begin{align}
\delta x(\nu)=&\frac{1}{\sqrt{2 \omega_0}}\l \delta a(\nu)+\delta a^{\dagger}(-\nu)\r,\\
\mathcal{F}_x(\nu)=&\frac{1}{\sqrt{2 \omega_0}}\bigg[\mathcal{F}_a(\nu)r(\nu)+\mathcal{F}_{a^{\dagger}}(-\nu)r^{*}(-\nu)\nonumber\\
&+\mathcal{F}^{+}(\nu)p(\nu)+\mathcal{F}^{-}(-\nu)p^{*}(-\nu)\bigg].\nonumber
\end{align}
Here, $\mathcal{F}_a\equiv\sqrt{2\kappa} a_{\text{in}}$ and the fluctuation operators $\delta a(\nu),\delta a^{\dagger}(-\nu)$ are defined in terms of noise operators in Eqs.~\eqref{EQ:PhotonComponent1}, \eqref{EQ:PhotonComponent2}. We evaluate them at the critical point, which is approached from the normal phase ($\braket{a}=\braket{\sigma^+}=0$, $\braket{\sigma^z}=-1$).
The complex functions $r(\nu)$ and $p(\nu)$ are defined as
\begin{align}
r(\nu)&=(\kappa-i (\nu+\omega_0))\label{EQ:r},\\
p(\nu)&=-\frac{4 g (\gamma-2i(\Delta+\nu))}{3\Delta^2+(\gamma-2i\nu)}\omega_0.\label{EQ:p}
\end{align}
In the $\delta x$-channel, the response of the fluctuations to the "driving force" $\mathcal{F}_{x}(\nu)$ is described by the equation
\begin{align}
\mathcal{F}_x(\nu)=&\l \omega^2_0-\frac{16 \Delta  g ^2}{4 \Delta ^2+(\gamma -2 i \nu )^2}\omega_0+(\kappa -i \nu )^2\r\delta x(\nu).
\label{EQ:HeisenbergLangevinEquationPhotoncomponent}
\end{align}
At low frequencies, equation Eq.\,\eqref{EQ:HeisenbergLangevinEquationPhotoncomponent} resembles a Langevin equation for a classical particle subject to a harmonic potential with oscillation frequency
\begin{align}
\alpha^2=\left(\omega^2_0 -\frac{16 \Delta  g ^2}{\gamma ^2+4 \Delta ^2}\omega_0+\kappa ^2\right)
\end{align}
and an effective damping constant
\begin{align}
\tilde{\kappa}= 2\left(\kappa+\frac{32 \gamma  \Delta  g ^2 \omega_0}{\left(\gamma ^2+4 \Delta ^2\right)^2}\right)\;.
\end{align}
This illustrates the fact that the photon can decay via two channels: directly via $\kappa$ 
and by first converting it to an atomic excitation, which can then decay via $\gamma$.

We may now identify the effective temperature of the system at the critical point as
\begin{align}
2 \tilde{\kappa}T^{\rm crit}_{\rm eff}&=\left.\lim\limits_{\nu \to 0}\frac{1}{2}\braket{\mathcal{F}_x(\nu)\mathcal{F}_x(-\nu)+\mathcal{F}_x(-\nu)\mathcal{F}_x(\nu)}\right|_{g=g_c},\;\\
T^{\rm crit}_{\rm eff}&=\frac{\left(\gamma ^2+4 \Delta ^2\right) \left(\kappa ^2+\omega_0^2\right) (\gamma  \omega_0+2 \Delta  \kappa )}{8 \Delta  \omega_0 \left(\gamma ^2 \kappa +2 \gamma  \left(\kappa ^2+\omega_0^2\right)+4 \Delta ^2 \kappa \right)}.\label{EQ:CritTemperature}
\end{align}
In the presence of only a single photonic decay channel, we recover the known cases \cite{dalla13}
\begin{align}
\lim\limits_{\gamma \to 0}T^{\rm crit}_{\rm eff}&=\frac{\kappa ^2+\omega_0^2}{4 \omega_0},\\
\lim\limits_{\kappa \to 0}T^{\rm crit}_{\rm eff}&=\frac{\frac{\gamma ^2}{4}+\Delta ^2}{4 \Delta }\;.
\end{align}
For certain parameter regimes $(\omega_0,\Delta,k,\gamma)>0$ the effective temperature in the presence of spontaneous decay $\gamma>0$ can be smaller than in the absence of atomic decay $\gamma=0$, i.e. $T^{\rm crit}_{\rm eff}< \lim\limits_{\gamma \to 0}T^{\rm crit}_{\rm eff}$.

This happens for $0<\Delta  <\frac{\kappa ^2+\omega_0^2}{\omega_0}$ and $0<\gamma <2 \sqrt{\frac{\Delta  \left(\omega_0 (\omega_0-\Delta )+\kappa ^2\right)}{\omega_0}}$, i.e. for system parameters for which spontaneous atomic decay is energetically favorable over cavity photon loss.

\section{Conclusions}
\label{subsec:conclu}

In this paper we investigated the effect of atomic spontaneous emission on the non-equilibrium steady-states of the open Dicke model. We argued that a site decoupling mean-field ansatz gives access to the exact solution 
of the problem in a long-time, thermodynamic limit. By determining the critical coupling $g_c(\kappa, \gamma, U)$ for the onset of superradiance as an explicit function of the spontaneous emission rate $\gamma$, we were able to compare this result to experimental 
values for the onset of superradiance as measured by Baden {\it et al.} \cite{baden14}. 
 Thereby we estimated an upper and lower bound for an effective spontaneous emission rate that might 
explain the experimentally observed 
discrepancy between previous analytical calculations and experimental measurements. 
Moreover, we have quantified the sideband asymmetry in the 
cavity-output spectrum due to atomic spontaneous emission.

An interesting future direction is the inclusion of additional short-range 
interactions between the atoms, for example by weakly dressing 
the spin-up level with a Rydberg state \cite{zeiher16}. This interaction will now compete
with cavity-mediated, long-range interactions and the various drive and 
decay processes. Moreover, changing the lattice geometry and using
space-dependent pump fields could enable synthesis of exotic, open 
quantum magnets whose properties are shaped by quantum fluctuations 
of the light field.

\acknowledgments
We thank M. Barrett for insightful explanations of experimental data. 
We also thank S. Diehl and A. Rosch for helpful discussions. 
This work was supported by the Leibniz Prize of A. Rosch and by the German Research Foundation (DFG) 
through the Institutional Strategy of the University of Cologne within the 
German Excellence Initiative (ZUK 81).
\newpage
\appendix
\section{Details for the effective temperature calculation}

We detail the calculation of the effective temperature in Sec.\,\ref{subsec:Teff} for the Dicke-phase transition in the presence of single-site spontaneous atomic decay.
The stochastic force operator satisfies the relation $\bigg[\mathcal{F}_x(-\nu)\bigg]^{\dagger}=\mathcal{F}_x(\nu)$ and obeys the commutation relation:
\begin{align}
&\frac{1}{2}\braket{\mathcal{F}_x(\nu)\mathcal{F}_x(\nu')+\mathcal{F}_x(\nu')\mathcal{F}_x(\nu)}=\nonumber \\
&\delta(\nu+\nu')\left(\frac{2\kappa}{4\omega_0}\bigg[r(\nu)r^{\dagger}(-\nu')+r(\nu')r^{\dagger}(-\nu)\bigg]\right. \nonumber\\
&\left. +\frac{\gamma}{4\omega_0}\bigg[p(\nu)p^{\dagger}(-\nu')+p(\nu')p^{\dagger}(-\nu)\bigg]\right)\;
\label{EQ:CommutationRelationPhoton}
\end{align}
Where $p(\nu)$ and $r(\nu)$ are given by Eq.\,\eqref{EQ:r} and by Eq.\,\eqref{EQ:p}.
At low frequencies the right-hand side of Eq.~(\ref{EQ:HeisenbergLangevinEquationPhotoncomponent}) evaluates to
\begin{align}
\left(\omega^2_0 -\frac{16 \Delta  g ^2}{\gamma ^2+4 \Delta ^2}\omega_0+\kappa ^2\right)-2i\nu  \left(\kappa+\frac{32 \gamma  \Delta  g ^2 \omega_0}{\left(\gamma ^2+4 \Delta ^2\right)^2}\right)
\end{align}
where we have dropped contributions $\mathcal{O}(\nu^2)$ and the commutation relation at $\nu'=-\nu$, see Eq.\,\eqref{EQ:CommutationRelationPhoton}, evaluates to: 
\begin{align}
&\frac{1}{2}\braket{\mathcal{F}_x(\nu)\mathcal{F}_x(\nu')+\mathcal{F}_x(\nu')\mathcal{F}_x(\nu)}\approx\frac{\kappa}{\omega_0}\left(\kappa ^2+\omega^2_0\right)\nonumber \\
&+\frac{\gamma}{4 \omega_0}\frac{32 g ^2}{\gamma^2+4\Delta^2}\omega^2_0+\mathcal{O}(\nu^2)
\end{align}
From which the critical temperature in Eq.\eqref{EQ:CritTemperature} can be calculated.
\section{Sub- and superradiant transition rates}
\label{App:SubandSuperradiantDecay}
One can check that the dissipators in Eq.\,\eqref{EQ:LindbladOperatorgamma} can be mapped exactly to the wavefunction approach that was derived previously in \cite{Scully2015}. As an example, we calculate the decay rates of the single photon sub- and superradiant states 
\begin{align}
\ket{-}&\equiv \ket{S=N/2-1,m=-N/2+1},\\
\ket{+}&\equiv\ket{S=N/2,m=-N/2+1},
\end{align}
respectively.
Both states are defined as
\begin{align}
\ket{+}&=\frac{1}{\sqrt{N}}\sum_{\ell=1}^N\ket{j},\\
\ket{-}&=\frac{1}{\sqrt{N}}\bigg[\sum_{j=1}^{N/2}\ket{j}-\sum_{j'=1}^{N/2}\ket{j'+N/2}\bigg]
\end{align}
where $\ket{j}=\ket{\downarrow,\dots,\downarrow,\uparrow_j,\downarrow\dots,\downarrow}$ labels the position of the one-atom excitation, $S$ is the total spin component and $m=(N_{\uparrow}-N_{\downarrow})/2$. We make use of Eq. \eqref{EQ:HeisenbergDynamicswithLindblad} for the time-evolution of the sub- and superradiant states as,
\begin{align}
&\partial_t \left(\ket{\pm}\bra{\pm}\right)=\gamma(1-\alpha)\bigg[\sum_{\ell=1}^N\left(\sigma^{+}_{\ell}\ket{\pm}\bra{\pm}\sigma^{-}_{\ell}\right)
-\ket{\pm}\bra{\pm}\bigg]\nonumber
\\
&+\gamma\alpha \bigg[S^{+}\ket{\pm}\bra{\pm}S^{-}-\frac{1}{2} 2 (S(S+1)-m(m-1))\ket{\pm}\bra{\pm}\bigg]\nonumber \\
\end{align}
The rate of emission is proportional to the population in the respective state. We can thus extract the super $(+)$ and subradiant rates $(-)$ as
\begin{align}
I_{\pm}&=\gamma\bigg[1-\alpha+\alpha(S(S+1)-m(m-1))\bigg]\\
I_{+}&=\gamma\left(1+(N-1)\alpha\right)\\
I_{-}&=\gamma\left(1-\alpha\right)
\end{align} 
It can be seen that the decay rate of the superradiant state is collectively enhanced, while the subradiant state has a lower decay rate than a single atomic excitation $I_{\text{single}}=\gamma$. The strength of the collective contribution is bounded by $0\leq \alpha \leq 1$, see App.\,\ref{App:SecCollectiveAtomicDecay}. 

\section{Cavity spectra in the superradiant regime}
We detail the calculations performed in Sec.\ref{sec:CavityOutPut} to obtain the cavity spectrum $S(\nu)$ that is defined in Eq.\,\eqref{Eq:CavSpectrumRelation}. The fluctuations $\delta a(\nu),\delta a^{\dagger}(\nu)$ around the photon condensate are given as
\begin{align}
\delta a(\nu)&=a^{\dagger}_{\rm in}(-\nu)f(\nu)+a_{\rm in}(\nu)g(\nu)\nonumber\\
&+\mathcal{F^{-}}(-\nu)m(\nu)+\mathcal{F^{+}}(\nu)h(\nu)+\mathcal{F}^z(\nu)\ell(\nu)\label{EQ:PhotonComponent1}\\
\delta a^{\dagger}(-\nu)&=a^{+}(-\nu)_{\rm in}g^{\dagger}(-\nu)+a_{\rm in}(\nu)f^{\dagger}(-\nu)\nonumber \\
&+\mathcal{F}^{-}(-\nu)h^{\dagger}(-\nu)+\mathcal{F}^{+}(\nu)m^{\dagger}(-\nu)+\mathcal{F}_{z}(\nu)\ell^{\dagger}(-\nu)\label{EQ:PhotonComponent2}
\end{align}
By employing Eqs.~(\ref{EQ:CorrelationFunction}-\ref{EQ:CorrelationFunctionz}) we can identify the cavity-spectrum as
\begin{align}
S(\nu)=&f^{\dagger}(\nu)f(\nu)+\gamma
h^{\dagger}(\nu)h(\nu)+2\gamma (1+\braket{\sigma^z})\ell(\nu)\ell^{\dagger}(\nu)\nonumber\\
&+2\gamma \braket{\sigma^-}h^{\dagger}(\nu)\ell(\nu)+2\gamma\braket{\sigma^+}\ell^{\dagger}(\nu)h(\nu)
\end{align}
Where the functions $\ell(\nu),h(\nu),f(\nu)$ are 
\begin{widetext}
\begin{align}
\Omega(\nu)=&-i (\gamma -i \nu ) \left(4 \Delta ^2+(\gamma -2 i \nu )^2\right) \left(\omega_0^2+(\kappa -i \nu )^2\right)+32 \Delta  g^3 \omega_0 (\braket{a^{\dagger}}+\braket{a}) (\braket{\sigma^-}-\braket{\sigma^+})\nonumber\\
&+8 g^2 \left(-(\braket{a^{\dagger}}+\braket{a})^2 (2 \nu +i \gamma ) (\kappa -i \nu )^2-\omega_0^2 (\braket{a^{\dagger}}+\braket{a})^2 (2 \nu +i \gamma )-2 \Delta  \omega_0 \braket{\sigma^z} (\nu +i \gamma )\right)\nonumber\\
\Omega(\nu)f(\nu)=&-16 i \Delta  g^2 \kappa  \left(2 g (\braket{a^{\dagger}}+\braket{a}) (\braket{\sigma^-}-\braket{\sigma^+})+\braket{\sigma^z} (-\nu -i \gamma )\right)\\
\Omega(\nu)h(\nu)=&-2 \sqrt{2\kappa} g  (\kappa -i (\nu +\omega_0)) \left(8 g^2 (\braket{a^{\dagger}}+\braket{a})^2+(\gamma -i \nu ) (\gamma +2 i (\Delta -\nu ))\right)\\
\Omega(\nu)\ell(\nu)=&-8 \sqrt{2\kappa} \Delta  g^2  (\braket{a^{\dagger}}+\braket{a}) (\kappa -i (\nu +\omega_0))
\end{align}
\end{widetext}
The cavity spectrum for the superradiant case for $g>g_c$ is given by
\begin{widetext}
\begin{align}
S(\nu)=&\frac{s(\nu)}{\Omega(\nu)\Omega^{*}(\nu)}\label{Eq:CavSpectrumSR}\\
\Omega(\nu)=&-i (\gamma -i \nu ) \left(4 \Delta ^2+(\gamma -2 i \nu )^2\right) \left(\omega_0^2+(\kappa -i \nu )^2\right)+32 \Delta  g^3 \omega_0 (\braket{a^{\dagger}}+\braket{a}) (\braket{\sigma^-}-\braket{\sigma^+})\nonumber\\
&-8 i g^2 \left((\braket{a^{\dagger}}+\braket{a})^2 (\gamma -2 i \nu ) (\kappa -i \nu )^2+\omega_0^2 (\braket{a^{\dagger}}+\braket{a})^2 (\gamma -2 i \nu )+2 \Delta  \omega_0 \braket{\sigma^z} (\gamma -i \nu )\right)\\
s(\nu)=&8 g^2 \kappa  \left\{\gamma  \left(\gamma ^2+\nu ^2\right) \left[\gamma ^2+4 (\Delta -\nu )^2\right] \left[\kappa ^2+(\nu +\omega_0)^2\right]+64 g^4 (\braket{a^{\dagger}}+\braket{a})^2 \left\{(\braket{a}+\braket{a^{\dagger}})^2 \gamma  \left[\kappa ^2+(\nu +\omega_0)^2\right]\right.\right. \nonumber\\
&\left.\left.-2 \Delta ^2 \kappa  (\braket{\sigma^-}-\braket{\sigma^+})^2\right\}
+16 g^2\left\{(\braket{a^{\dagger}}+\braket{a})^2 \gamma \left[\gamma ^2+2 \left(\Delta ^2+\Delta  \nu -\nu ^2\right)\right] \left[\kappa ^2+(\nu +\omega_0)^2\right]+2 \Delta ^2 \kappa  \braket{\sigma^z}^2 \left(\gamma ^2+\nu ^2\right)\right.\right.\nonumber\\ 
&\left.+2 \gamma  \braket{\sigma^z} (\braket{a^{\dagger}}+\braket{a})^2 \Delta ^2\left[\kappa ^2+(\nu +\omega_0)^2\right]\right\}\nonumber\\ 
&\left.+64 \gamma  \Delta  g^3 (\braket{a^{\dagger}}+\braket{a}) \left[(\braket{a^{\dagger}}+\braket{a})^2 (\braket{\sigma^-}+\braket{\sigma^+}) \left[\kappa ^2+(\nu +\omega_0)^2\right]+2 i \Delta  \kappa  \braket{\sigma^z} (\braket{\sigma^-}-\braket{\sigma^+})\right]\right.\nonumber\\
&+\left.8 \gamma  \Delta  g (\braket{a^{\dagger}}+\braket{a}) \left[\kappa ^2+(\nu +\omega_0)^2\right] \left[\gamma ^2 (\braket{\sigma^-}+\braket{\sigma^+})-i \gamma  (2 \Delta -3 \nu ) (\braket{\sigma^-}-\braket{\sigma^+})+2 \nu  (\Delta -\nu ) (\braket{\sigma^-}+\braket{\sigma^+})\right]\right\}
\end{align}
\end{widetext}
At the phase transition, there are two poles that become purely imaginary and describe the over-damped dynamics. The corresponding expressions are obtained by expanding ${\rm det}[\b{G^{-1}_{R}}(\nu)]=0$ up to second-order in the frequency with $(\beta, U)\to 0$. We refer to the solutions of the resulting quadratic equation by $(\nu_1,\nu_2)$. 
The first pole vanishes linearly in $(g-g_c)$ and is given as:
\begin{align}
\nu_{1}=\frac{8 i \Delta  \omega_0 \left(\gamma ^2+4 \Delta ^2\right) (g -g_c) (g +g_c)}{\kappa  \left(\gamma ^2+4 \Delta ^2\right)^2+32 \gamma  \Delta  g_c^2 \omega_0}\label{EQ:CriticalPole1}
\end{align}
The residual pole at $g=g_c$ is given by Eq.\,\eqref{EQ:GammaPolariton}, with the limit
\begin{align}
\lim_{\kappa \to 0}\nu_2&=- i \gamma  \frac{\omega_0^2}{\frac{\gamma ^2}{4}+ \left(\Delta ^2+\omega_0^2\right)}\label{EQ:CriticalPole2}
\end{align}

\section{Heisenberg-Langevin-Framework}
\label{App:MakrovBath}
We review the calculation for the set of correlation functions for the noise-operators given in Eq.\,\eqref{EQ:NoiseOperatorsCorrelsAtom}. The general scheme for the derivation of noise-operators and their correlations is outlined for instance in Ref \cite{scully}. For this we apply the standard Heisenberg-Langevin theory where the interaction of the system with the external bath is specified in terms of a Hamiltonian that couples the bath modes linearly to the system operators. The bath outside the cavity is the continuum of radiation modes.
We consider the system-bath interaction in the interaction picture
\begin{align}
H^{\rm atoms}_{\rm bath-sys}(t)&=\sum_{k,\ell=1}^N \left(g_{k,\ell}\sigma^+_{\ell}b_k e^{i(\Delta-\nu_k)t}+cc.\right)\\
H^{\rm photons}_{\rm bath-sys}(t)&=\sum_{k} \left(\tilde{g}_k a^{\dagger}c_k e^{i(\omega_0-\omega_k)t}+cc.\right)
\end{align}
The equation of motions for the system and the bath operators can be written as
\begin{align}
\partial_t a_t&=-i[a_t,H_t]=-i\sum_k \tilde{g}_k c_{k,t}e^{i(\omega_0-\omega_k)t}\label{EQ:SystemBath0}\\
\partial_t b_{k,t}&=-i[b_{k,t},H_t]=-i\sum_{\ell=1}^N g^{*}_{k,\ell}\sigma^{-}_{\ell,t}e^{-i(\Delta-\nu_k)t}\\
\partial_t c_{k,t}&=-i[b_{k,t},H_t]=-i\sum_{\ell=1}^N \tilde{g}^{*}_{k,\ell}a_t e^{-i(\Delta-\omega_k)t}\\
\partial \sigma^{-}_{\ell',t}&=-i[\sigma^{-}_{\ell',t},H_t]=i\sum_{k}\sigma^{z}_{\ell',t}b_{k,t}e^{i(\Delta-\nu_k)t}g_{k,\ell}\label{EQ:SystemBath1}\\
\partial \sigma^{z}_{\ell',t}&=-i[\sigma^{z}_{\ell,t},H_t]\nonumber \\
&=\sum_k\left(-2i\sigma^{+}_{\ell',t}b_{k,t}e^{i(\Delta-\nu_k)t}g_{k,\ell'} -cc.\right)\label{EQ:SystemBath2}.
\end{align}
Here we have used $H_t=H^{\rm atoms}_{\rm bath-sys}(t)+H^{\rm photons}_{\rm bath-sys}(t)$. We integrate the equations of the bath modes
\begin{align}
b_{k,t}=b_{k,0}-i\int_{0}^t dt' \sum_{\ell=1}^N g^{*}_{k,\ell}\sigma^{-}_{\ell,t'}e^{-i(\Delta-\nu_k)t'}\label{EQ:BathModeAtom}\\
c_{k,t}=c_{k,0}-i\int_{0}^t dt' \sum_{\ell=1}^N \tilde{g}^{*}_{k,\ell}a_{t'}e^{-i(\Delta-\omega_k)t'}\label{EQ:BathModePhoton}
\end{align}
and insert Eqs.\,(\ref{EQ:BathModeAtom}-\ref{EQ:BathModePhoton}) and the conjugates into Eqs.\,(\ref{EQ:SystemBath0}-\ref{EQ:SystemBath2}).
\begin{align}
&\partial_t a_t=\mathcal{F}_{a_t}-\int_{0}^t dt' \sum_k |\tilde{g}_k|^2 \xi^{*}_{k,\omega_0,t,t'}a_{t'}\\
&\partial_t \sigma^{+}_{\ell',t}=\mathcal{F}^{+}_{\ell',t}\nonumber\\
&+\int_{0}^t dt' \sum_{\ell,k} |g_k|^2 e^{i(\b{k}-\b{k_0})(\b{r_{\ell}}-\b{r_{\ell'}})}\xi_{k,\Delta,t,t'}\sigma^{+}_{\ell,t'}\sigma^z_{\ell',t}\\
&\partial_t \sigma^{-}_{\ell',t}=\mathcal{F}^{-}_{\ell',t}\nonumber\\
&+\int_{0}^t dt' \sum_{\ell,k} |g_k|^2 e^{-i(\b{k}-\b{k_0})(\b{r_{\ell}}-\b{r_{\ell'}})}\xi^{*}_{k,\Delta,t,t'}\sigma^z_{\ell',t}\sigma^{-}_{\ell,t'}\\
&\partial_t \sigma^z_{\ell',t}=\mathcal{F}^z_{\ell',t}\nonumber \\
&-2\left(\int_{0}^t dt' \sigma^{+}_{\ell',t}\sum_{\ell}|g_k|^2 e^{i(\b{k}-\b{k_0})(\b{r_{\ell}}-\b{r_{\ell'}})} \xi_{k,\Delta,t,t'} \sigma^{-}_{\ell,t'}+cc.\right)
\end{align}
The noise-operators are given by 
\begin{align}
\mathcal{F}_{a_t}&=-i\sum_k \tilde{g}_k c_k(0)e^{-i(\omega_0-\omega_k)t}\\
\mathcal{F}^{+}_{\ell',t}&=-i \sum_{k}b^{+}_{k}(0)\sigma^z_{\ell'}(t)e^{-i(\Delta-\nu_k)t}g^{*}_k e^{i(\b{k}-\b{k_0})\b{r}_{\ell'}}\label{EQ:NoiseAtomsx}\\
\mathcal{F}^{-}_{\ell',t}&=i \sum_{k}\sigma^z_{\ell',t}b_{k}(0)e^{i(\Delta-\nu_k)t}g_k e^{-i(\b{k}-\b{k_0})\b{r}_{\ell'}}\\
\mathcal{F}^{z}_{\ell ',t}&=\sum_{k}\left(2i b^{+}_{k}(0)\sigma^{-}_{\ell',t}g^{*}_k e^{-i(\Delta-\nu_k)t}e^{i(\b{k}-\b{k_0})\b{r}_{\ell'}}+ cc.\right)\label{EQ:NoiseAtomsz}
\end{align}
with the following relations
\begin{align}
\frac{\gamma}{2}\delta(t-t')&=\sum_k |g_k|^2 \xi_{k,t,t'}=2\pi |g_{\Delta}|^2\mathcal{D}(\Delta)\delta(t-t'),\\
\kappa\delta(t-t')&=\sum_k |\tilde{g}_k|^2 \xi_k(t,t')=2\pi |g_{\omega_0}|^2\mathcal{D}(\omega_0)\delta(t-t'),\\
\xi_{k,\omega,t,t'}&=\exp\left(-i(\omega-\nu_k)(t-t')\right).
\end{align}
Here, we have taken $g_{k,\ell}=g_k e^{-i(\b{k}-\b{k}_0)\b{r}_{\ell}}$ as the cavity-shifted, spatially dependent atom-photon coupling to the bath modes $b_k$ outside of the cavity, where $\b{k_0}$ is the cavity wave vector. The strength of the collective decay component is derived in App.\,\ref{App:SecCollectiveAtomicDecay}.
We finally obtain the Heisenberg-Langevin equations that result from the interaction with the bath. 
\begin{align}
\partial_t a_t&=\mathcal{F}_{a_t}-\kappa a_t\\
\partial_t \sigma^{+}_{\ell',t}&=\mathcal{F}^{+}_{\ell',t}-\frac{\gamma}{2}\sigma^{+}_{\ell',t}+\frac{\gamma \alpha}{2}\sum_{\ell\neq \ell'} \sigma^{+}_{\ell}(t)\sigma^{z}_{\ell',t}\\
\partial_t \sigma^{-}_{\ell',t}&=\mathcal{F}^{-}_{\ell',t}-\frac{\gamma}{2}\sigma^{-}_{\ell',t}+\frac{\gamma \alpha}{2}\sum_{\ell\neq \ell'} \sigma^{z}_{\ell',t}\sigma^{-}_{\ell,t}\\
\partial_t \sigma^z_{\ell',t}&=\mathcal{F}^z_{\ell',t}-(1+\sigma^{z}_{\ell',t})\gamma-\gamma \alpha \sum_{\ell\neq \ell'}\left(\sigma^{+}_{\ell',t}\sigma^{-}_{\ell,t}+cc.\right)
\end{align}
The second moment of the Noise correlations function can be evaluated with the help of Eqs.~(\ref{EQ:NoiseAtomsx}-\ref{EQ:NoiseAtomsz}) leading to the noise correlation matrix in Eq.\,\eqref{EQ:NoiseOperatorsCorrelsAtom}. For the photonic noise operators we find $\braket{\mathcal{F}_{a_t}\mathcal{F}_{a^\dagger_t}}=2\kappa \delta(t-t')$ as the only non-vanishing correlator at the zero-temperature bath, see also Eq.\,\eqref{EQ:NoiseCorrelator2} where we have redefined the noise-operator according to $\mathcal{F}_{a}(t)=\sqrt{2\kappa}a_{\rm in,t}$.
\onecolumngrid
\section{Non-local Lindblad contribution in Born-Markov approximation}
\label{App:SecCollectiveAtomicDecay}
For the derivation of the collective decay contribution, we will briefly review the textbook approach (see e.g.\,\cite{Steck}) of how the external reservoir influences the evolution of the system in a Born-Markov approximation. This leads to a in general non-local density matrix equation, see \eqref{EQ:GeneralLindbladNonLocalDensityMatrix}. Collective decay contributions that add to the single atom decay rates have been derived in the context of single photon sub- and superradiant states \cite{Scully2015} in a wave-function formalism. Here, we carry these considerations over to a density matrix formalism and show that a description in terms of Lindblad operators reproduces the results obtained from the wavefunction picture. The collective loss contribution emerges by allowing all spins to interact with one shared bath. 
In the interaction picture, the system bath Hamiltonian in the rotating wave approximation can be written as

\begin{align}
H^{\rm atoms}_{\rm bath-sys}(t)&=\sum_{k}\sum_{\ell=1}^N \left(g_{k,\ell}\sigma^+_{\ell}b_k e^{i(\Delta-\nu_k)t}+g^{*}_{k,\ell} b^{\dagger}_k\sigma^{-}_{\ell} e^{-i(\Delta-\nu_k)t}\right).
\end{align}
Here, the spatially dependent coupling to the bath is given by $g_{k,\ell}=g_k e^{-i \b{k} \b{r}_{\ell}}$. The time-evolution of the full system and bath density matrix that is generated by the system-bath Hamiltonian reads
\begin{align}
\partial_t \rho(t)_{\rm}&=-\frac{i}{\hbar}\bigg[H^{\rm atoms}_{\rm bath-sys}(t),\rho(0)-\frac{i}{\hbar}\int_{0}^t dt' \bigg[H^{\rm atoms}_{\rm bath-sys}(t'),\rho(t')\bigg]\bigg].
\label{EQ:DensityMatrix}
\end{align}
With a weak system reservoir coupling, the perturbative Ansatz for the density matrix is written as
\begin{align} 
\rho(t')\equiv \rho_{\rm bath-sys}(t')\approx \rho_{\rm sys}(t')\otimes \rho_{\rm bath}(0)+\delta \rho_{\rm bath-sys}(t'),
\end{align}
where the last term is of order $\mathcal{O}(g_k)$ and the Born approximation corresponds to assuming that the time evolution of the bath is unaffected by the time-evolution of the system, $\rho_{\rm bath}(t')\approx \rho(0)$. These approximations and expansions are justified by assuming a large reservoir that has vanishing correlations with the system. The Markov approximation corresponds to assuming that the evolution of the system is memoryless $\rho(t')\rightarrow \rho(t)$.  Additionally, we phase shift the atomic operators around the cavity resonance. This accounts for the fact that the cavity mode couples to the atoms with a position dependent phase factor characterised by the cavity wavevector
\begin{align}
\sigma^{+}_{\ell} \to \sigma^{+}_{\ell}e^{-i \b{k}_{0}\b{r}_{\ell}}, \quad \sigma^{-}_{\ell} \to \sigma^{-}_{\ell} e^{i \b{k}_0 \b{r}_{\ell}}.
\end{align}

After tracing out all bath degrees of freedom, we arrive at the non-local density matrix equation for the system only,
\begin{align}
\partial_t \rho_{\rm sys}(t)=&\frac{1}{\hbar^2}\int_0^t dt' \sum_{k,\ell,\ell'}|g_{k}|^2 e^{-i(\b{k}-\b{k}_0)(\b{r}_{\ell'}-\b{r}_{\ell})} \bigg[\sigma^{-}_{\ell'}\rho_t\sigma^{+}_{\ell}\left(\zeta_{k}+\zeta^{*}_{k}\right)-
\rho_t\sigma^{+}_{\ell'}\sigma^{-}_{\ell}\zeta^{*}_{k}-
\sigma^{+}_{\ell'}\sigma^{-}_{\ell}\rho_t\zeta_{k}\bigg].\label{EQ:GeneralLindbladNonLocalDensityMatrix}
\end{align}
We have collected temporal phase factors as $\zeta_{k}(t'-t)=\exp\left(-i(\Delta-\nu_k)(t'-t)\right)$. This equation has two important contributions. The first describes the local dissipators obtained from the $\ell=\ell'$ terms. The second contribution is of collective nature and obtained from the condition $|\b{k}-\b{k}_0|\approx 0$. We note that contributions with $\ell\neq\ell'$ are generally suppressed by a factor $\propto 1/V$.
We single out the one-atom loss contribution
\begin{align}
\partial_t \rho_{\rm sys}(t)=&\gamma\sum_{\ell=1}^N \left(\sigma^{-}_{\ell}\rho_t\sigma^{+}_{\ell}-\frac{1}{2}\{\sigma^{+}_{\ell}\sigma^{-}_{\ell},\rho_t\}\right)\nonumber\\
&+\int_0^t dt' \sum_{k,\ell'\neq\ell}e^{-i(\b{k}-\b{k}_0)(\b{r}_{\ell'}-\b{r}_{\ell})}|g_{k}|^2\left( \sigma^{-}_{\ell'}\rho_t\sigma^{+}_{\ell}\left(\zeta_{k}+\zeta^{*}_{k}\right)-\rho_t\sigma^{+}_{\ell'}\sigma^{-}_{\ell}\zeta^{*}_{k}-\sigma^{+}_{\ell'}\sigma^{-}_{\ell}\rho_t\zeta_{k}\right)\label{EQ:NonLocalLindbladContribution}
\end{align}   
Focusing only on the collective contribution to the Lindblad equation that arises from wave vectors $|\b{k}-\b{k}_0|\approx 0$, we proceed by calculating the weight of the associated delta function
\begin{align}
&\int_0^t dt' \sum_{k} |g_k|^2 e^{-i(\b{k}-\b{k}_0)(\b{r}_{\ell'}-\b{r}_{\ell})}\zeta_k \approx \int_0^t dt' \sum_{k}|g_k|^2 \frac{(2\pi)^3}{V}\delta(\b{k}-\b{k}_0)\zeta_k
\end{align}
We assume that the bath modes lie dense and work in the continuum limit to use the replacements
\begin{align}
\delta(\b{k}-\b{k}_0)&=\frac{1}{2\pi}\int_{-R}^R e^{i(k-k_0)r}\delta(\theta_k-\theta_{k_0})\delta(\phi_k-\phi_{k_0})dr\frac{1}{k^2 \sin(\theta_k)}\\
\sum_k &\to \frac{V}{(2\pi)^3}\int_0^{\infty} dk k^2 \int_{0}^{\pi}\sin(\theta_k) d\theta_k \int_0^{2\pi} d\phi_k.
\end{align}
which leads to the integral
\begin{align}
\int_0^t dt' \sum_{k} |g_k|^2 e^{-i(\b{k}-\b{k}_0)(\b{r}_{\ell'}-\b{r}_{\ell})}\zeta_k \approx& \int_0^t dt' \int_0^\infty dk |g_k|^2 \bigg[\frac{1}{2\pi}\int_{-R}^R e^{-i(k-k_0)r}dr\bigg] e^{i(ck-\Delta)(t-t')}\nonumber\\
&= \int_0^t dt' \int_0^\infty dk |g_k|^2 \frac{1}{2\pi}\int_{-R}^R \exp\bigg[ic(k-k_0)(t-t'-r/c)+i c(k_0-k_{\Delta})(t-t')\bigg] dr \nonumber \\
&= |g_{k_0}|^2\int_0^t dt'\int_{-R}^R dr \frac{1}{2\pi}\frac{2\pi}{c}\delta(t-t'-r/c)\exp\bigg[i c(k_0-k_{\Delta})(t-t')\bigg]\nonumber\\
&=|g_{k_0}|^2 \int_{-R}^{R}dr \frac{1}{2c}\frac{R}{c}|g_{k_0}|^2 \exp\bigg[i c(k_0-k_{\Delta})r/c\bigg]\nonumber \\
&=|g_{k_0}|^2\frac{\sin\left((k_0-k_{\Delta})R\right)}{c(k_0-k_{\Delta})}
\end{align}

Here $R$ is the radius of the atomic cloud in the cavity which is much larger than the cavity wavelength. 
In the last line we have used that $|g_k|^2$ does not vary significantly around $k\sim k_0$ and pull it out of the integral
\begin{align}
\int_0^{\infty} dk |g_{k}|^2\exp\bigg[ic(k-k_0)(t-t'-r/c)\bigg]&=\frac{2\pi}{c}\delta(t-t'-r/c)|g_{k_0}|^2.
\end{align}
When the system becomes superradiant in the steady-state, the direction of emission is locked to the cavity characterised by its wavenumber $k_0$. Taking the difference of the cavity wavenumber ($k_0$) and the wavenumber corresponding to the atomic resonance ($k_{\Delta}=\Delta/c$) as small, one can expand $\sin\left((k_0-k_{\Delta})R\right)/c(k_0-k_{\Delta})\sim R/c$. 

\begin{align}
\int_0^t dt' \sum_{k} |g_k|^2 e^{i(\b{k}-\b{k}_0)(\b{r}_{\ell}-\b{r}_{\ell'})}\zeta_k &= \frac{\sin\left((k_0-k_{\Delta})R\right)}{c(k_0-k_{\Delta})}|g_{k_0}|^2 \approx \frac{2\pi D(k_{\Delta})}{2\pi D(k_{\Delta})}\frac{R}{c}|g_{k_{\Delta}}|^2\nonumber\\
&=\frac{\gamma}{2}\frac{R}{c}\frac{1}{2\pi D(k_{\Delta})}=\frac{\gamma}{2}\frac{3}{8\pi}\left(\frac{\lambda_{\Delta}^2}{4\pi R^2}\right)\equiv \frac{\gamma}{2}\alpha
\label{Eq:AlphaCollectiveContribution}
\end{align}
here, the volume of the atomic sample is taken to be $V=(4/3) \pi R^3$ and the density of states $D(k_{\Delta})=Vk_{\Delta}^2/\pi^2 c$. The strength for the collective decay that is here calculated in a Lindblad formalism is equivalent to a wavefunction picture obtained in \cite{Scully2015}. 
The geometric factor for the collective decay is then given by
\begin{align}
\lim_{k_{\Delta} \to k_0}\alpha=\frac{3}{8\pi}\left(\frac{\lambda_0^2}{4\pi R^2}\right)
\label{EQ:geometricfactor}
\end{align} 
Eq.\,\eqref{Eq:AlphaCollectiveContribution} and Eq.\,\eqref{EQ:geometricfactor} determine the strength of collective losses in a large sample limit $R\gg \lambda_0$. 
In the above limit, it can be seen that the collective decay is weak. If $k_{\Delta}$ and $k_0$ deviate from another we expect this suppression to be even stronger as pointed out in Sec. \ref{subsec:langevin}. Since we do not have the exact microscopic coupling constants we interpret our conservative estimate as an upper bound for the strength of the collective decay. 
Using Eq.\,\eqref{Eq:AlphaCollectiveContribution} in Eq.\,\eqref{EQ:NonLocalLindbladContribution} leads to 
\begin{align}
\partial_t \rho_{\rm sys}(t)=&\gamma\sum_{\ell=1}^N \left(\sigma^{-}_{\ell}\rho_t\sigma^{+}_{\ell}-\frac{1}{2}\{\sigma^{+}_{\ell}\sigma^{-}_{\ell},\rho_t\}\right)+\gamma \alpha \sum_{\ell'\neq\ell}\left(\sigma^{-}_{\ell'}\rho_t\sigma^{+}_{\ell}-\frac{1}{2}\{\sigma^{+}_{\ell'}\sigma^{-}_{\ell},\rho_t\}\right)
\label{EQ:LocalandNonLocalDissipators}
\end{align}
It is possible to roughly estimate the order of magnitude of the geometric factor $\alpha$ that determines the collective loss rates. 
For the experimental realisation of the super radiance transition with cavity-assisted Raman transitions, an intracavity optical lattice was used to trap the atoms such that they all couple with the same strength to the cavity mode, i.e.\, the optical lattice was chosen to be commensurate with the cavity mode function. For the estimation of the geometric coupling strength, we take the cavity parameters from the Singapore setup as detailed in \cite{Barrett2012} and in \cite{baden14} with Eq.\,\eqref{EQ:geometricfactor}. 
\begin{align}
\alpha \sim\frac{3}{32\pi^2}\left(\frac{39}{50000}/(1/10)\right)^2 \sim 6 \times 10^{-7}
\end{align}
The collective emission rates appearing in the mean-field equations are enhanced by the number of atoms loaded into the lattice which we take as $N\sim10^4$ which would leave us with a conservative estimate of $\beta=\alpha N\sim 10^{-3}$.
\twocolumngrid
\twocolumngrid
\bibliographystyle{apsrev4-1_custom}
\bibliography{20171705_PhysRevAResubmit}

\begin{thebibliography}{33}%
\makeatletter
\providecommand \@ifxundefined [1]{%
 \@ifx{#1\undefined}
}%
\providecommand \@ifnum [1]{%
 \ifnum #1\expandafter \@firstoftwo
 \else \expandafter \@secondoftwo
 \fi
}%
\providecommand \@ifx [1]{%
 \ifx #1\expandafter \@firstoftwo
 \else \expandafter \@secondoftwo
 \fi
}%
\providecommand \natexlab [1]{#1}%
\providecommand \enquote  [1]{``#1''}%
\providecommand \bibnamefont  [1]{#1}%
\providecommand \bibfnamefont [1]{#1}%
\providecommand \citenamefont [1]{#1}%
\providecommand \href@noop [0]{\@secondoftwo}%
\providecommand \href [0]{\begingroup \@sanitize@url \@href}%
\providecommand \@href[1]{\@@startlink{#1}\@@href}%
\providecommand \@@href[1]{\endgroup#1\@@endlink}%
\providecommand \@sanitize@url [0]{\catcode `\\12\catcode `\$12\catcode
  `\&12\catcode `\#12\catcode `\^12\catcode `\_12\catcode `\%12\relax}%
\providecommand \@@startlink[1]{}%
\providecommand \@@endlink[0]{}%
\providecommand \url  [0]{\begingroup\@sanitize@url \@url }%
\providecommand \@url [1]{\endgroup\@href {#1}{\urlprefix }}%
\providecommand \urlprefix  [0]{URL }%
\providecommand \Eprint [0]{\href }%
\providecommand \doibase [0]{http://dx.doi.org/}%
\providecommand \selectlanguage [0]{\@gobble}%
\providecommand \bibinfo  [0]{\@secondoftwo}%
\providecommand \bibfield  [0]{\@secondoftwo}%
\providecommand \translation [1]{[#1]}%
\providecommand \BibitemOpen [0]{}%
\providecommand \bibitemStop [0]{}%
\providecommand \bibitemNoStop [0]{.\EOS\space}%
\providecommand \EOS [0]{\spacefactor3000\relax}%
\providecommand \BibitemShut  [1]{\csname bibitem#1\endcsname}%
\let\auto@bib@innerbib\@empty
\bibitem [{\citenamefont {Gardiner}\ and\ \citenamefont
  {Zoller}(2015)}]{gardiner15}%
  \BibitemOpen
  \bibfield  {author} {\bibinfo {author} {\bibfnamefont {C.}~\bibnamefont
  {Gardiner}}\ and\ \bibinfo {author} {\bibfnamefont {P.}~\bibnamefont
  {Zoller}},\ }\href
  {http://www.worldscientific.com/worldscibooks/10.1142/p983} {\emph {\bibinfo
  {title} {The Quantum World of Ultra-Cold Atoms and Light Book II: The Physics
  of Quantum-Optical Devices}}},\ \bibinfo {edition} {1st}\ ed.\ (\bibinfo
  {publisher} {World Scientific},\ \bibinfo {year} {2015})\BibitemShut
  {NoStop}%
\bibitem [{\citenamefont {Klinder}\ \emph {et~al.}(2015)\citenamefont
  {Klinder}, \citenamefont {Ke\ss{}ler}, \citenamefont {Bakhtiari},
  \citenamefont {Thorwart},\ and\ \citenamefont {Hemmerich}}]{klinder15}%
  \BibitemOpen
  \bibfield  {author} {\bibinfo {author} {\bibfnamefont {J.}~\bibnamefont
  {Klinder}}, \bibinfo {author} {\bibfnamefont {H.}~\bibnamefont {Ke\ss{}ler}},
  \bibinfo {author} {\bibfnamefont {M.~R.}\ \bibnamefont {Bakhtiari}}, \bibinfo
  {author} {\bibfnamefont {M.}~\bibnamefont {Thorwart}}, \ and\ \bibinfo
  {author} {\bibfnamefont {A.}~\bibnamefont {Hemmerich}},\ }\bibfield  {title}
  {\enquote {\bibinfo {title} {Observation of a superradiant mott insulator in
  the dicke-hubbard model},}\ }\href {\doibase 10.1103/PhysRevLett.115.230403}
  {\bibfield  {journal} {\bibinfo  {journal} {Phys. Rev. Lett.}\ }\textbf
  {\bibinfo {volume} {115}},\ \bibinfo {pages} {230403} (\bibinfo {year}
  {2015})}\BibitemShut {NoStop}%
\bibitem [{\citenamefont {Landig}\ \emph {et~al.}(2016)\citenamefont {Landig},
  \citenamefont {Hruby}, \citenamefont {Dogra}, \citenamefont {Landini},
  \citenamefont {Mottl}, \citenamefont {Donner},\ and\ \citenamefont
  {Esslinger}}]{landig16}%
  \BibitemOpen
  \bibfield  {author} {\bibinfo {author} {\bibfnamefont {R.}~\bibnamefont
  {Landig}}, \bibinfo {author} {\bibfnamefont {L.}~\bibnamefont {Hruby}},
  \bibinfo {author} {\bibfnamefont {N.}~\bibnamefont {Dogra}}, \bibinfo
  {author} {\bibfnamefont {M.}~\bibnamefont {Landini}}, \bibinfo {author}
  {\bibfnamefont {R.}~\bibnamefont {Mottl}}, \bibinfo {author} {\bibfnamefont
  {T.}~\bibnamefont {Donner}}, \ and\ \bibinfo {author} {\bibfnamefont
  {T.}~\bibnamefont {Esslinger}},\ }\bibfield  {title} {\enquote {\bibinfo
  {title} {Quantum phases from competing short- and long-range interactions in
  an optical lattice},}\ }\href {http://dx.doi.org/10.1038/nature17409}
  {\bibfield  {journal} {\bibinfo  {journal} {Nature}\ }\textbf {\bibinfo
  {volume} {532}},\ \bibinfo {pages} {476} (\bibinfo {year}
  {2016})}\BibitemShut {NoStop}%
\bibitem [{\citenamefont {Neuzner}\ \emph {et~al.}(2016)\citenamefont
  {Neuzner}, \citenamefont {K{\"o}rber}, \citenamefont {Morin}, \citenamefont
  {Ritter},\ and\ \citenamefont {Rempe}}]{neuzner16}%
  \BibitemOpen
  \bibfield  {author} {\bibinfo {author} {\bibfnamefont {A.}~\bibnamefont
  {Neuzner}}, \bibinfo {author} {\bibfnamefont {M.}~\bibnamefont {K{\"o}rber}},
  \bibinfo {author} {\bibfnamefont {O.}~\bibnamefont {Morin}}, \bibinfo
  {author} {\bibfnamefont {S.}~\bibnamefont {Ritter}}, \ and\ \bibinfo {author}
  {\bibfnamefont {G.}~\bibnamefont {Rempe}},\ }\bibfield  {title} {\enquote
  {\bibinfo {title} {Interference and dynamics of light from a
  distance-controlled atom pair in an optical cavity},}\ }\href
  {http://dx.doi.org/10.1038/nphoton.2016.19} {\bibfield  {journal} {\bibinfo
  {journal} {Nat Photon}\ }\textbf {\bibinfo {volume} {10}},\ \bibinfo {pages}
  {303} (\bibinfo {year} {2016})}\BibitemShut {NoStop}%
\bibitem [{\citenamefont {Strack}\ and\ \citenamefont
  {Sachdev}(2011)}]{strack11}%
  \BibitemOpen
  \bibfield  {author} {\bibinfo {author} {\bibfnamefont {P.}~\bibnamefont
  {Strack}}\ and\ \bibinfo {author} {\bibfnamefont {S.}~\bibnamefont
  {Sachdev}},\ }\bibfield  {title} {\enquote {\bibinfo {title} {Dicke quantum
  spin glass of atoms and photons},}\ }\href {\doibase
  10.1103/PhysRevLett.107.277202} {\bibfield  {journal} {\bibinfo  {journal}
  {Phys. Rev. Lett.}\ }\textbf {\bibinfo {volume} {107}},\ \bibinfo {pages}
  {277202} (\bibinfo {year} {2011})}\BibitemShut {NoStop}%
\bibitem [{\citenamefont {Gopalakrishnan}\ \emph {et~al.}(2011)\citenamefont
  {Gopalakrishnan}, \citenamefont {Lev},\ and\ \citenamefont
  {Goldbart}}]{sarang11}%
  \BibitemOpen
  \bibfield  {author} {\bibinfo {author} {\bibfnamefont {S.}~\bibnamefont
  {Gopalakrishnan}}, \bibinfo {author} {\bibfnamefont {B.~L.}\ \bibnamefont
  {Lev}}, \ and\ \bibinfo {author} {\bibfnamefont {P.~M.}\ \bibnamefont
  {Goldbart}},\ }\bibfield  {title} {\enquote {\bibinfo {title} {Frustration
  and glassiness in spin models with cavity-mediated interactions},}\ }\href
  {\doibase 10.1103/PhysRevLett.107.277201} {\bibfield  {journal} {\bibinfo
  {journal} {Phys. Rev. Lett.}\ }\textbf {\bibinfo {volume} {107}},\ \bibinfo
  {pages} {277201} (\bibinfo {year} {2011})}\BibitemShut {NoStop}%
\bibitem [{\citenamefont {Buchhold}\ \emph {et~al.}(2013)\citenamefont
  {Buchhold}, \citenamefont {Strack}, \citenamefont {Sachdev},\ and\
  \citenamefont {Diehl}}]{bunti13}%
  \BibitemOpen
  \bibfield  {author} {\bibinfo {author} {\bibfnamefont {M.}~\bibnamefont
  {Buchhold}}, \bibinfo {author} {\bibfnamefont {P.}~\bibnamefont {Strack}},
  \bibinfo {author} {\bibfnamefont {S.}~\bibnamefont {Sachdev}}, \ and\
  \bibinfo {author} {\bibfnamefont {S.}~\bibnamefont {Diehl}},\ }\bibfield
  {title} {\enquote {\bibinfo {title} {Dicke-model quantum spin and photon
  glass in optical cavities: Nonequilibrium theory and experimental
  signatures},}\ }\href {\doibase 10.1103/PhysRevA.87.063622} {\bibfield
  {journal} {\bibinfo  {journal} {Phys. Rev. A}\ }\textbf {\bibinfo {volume}
  {87}},\ \bibinfo {pages} {063622} (\bibinfo {year} {2013})}\BibitemShut
  {NoStop}%
\bibitem [{\citenamefont {Hepp}\ and\ \citenamefont {Lieb}(1973)}]{hepp73}%
  \BibitemOpen
  \bibfield  {author} {\bibinfo {author} {\bibfnamefont {K.}~\bibnamefont
  {Hepp}}\ and\ \bibinfo {author} {\bibfnamefont {E.~H.}\ \bibnamefont
  {Lieb}},\ }\bibfield  {title} {\enquote {\bibinfo {title} {On the
  superradiant phase transition for molecules in a quantized radiation field:
  the dicke maser model},}\ }\href {\doibase 10.1016/0003-4916(73)90039-0}
  {\bibfield  {journal} {\bibinfo  {journal} {Annals of Physics}\ }\textbf
  {\bibinfo {volume} {76}},\ \bibinfo {pages} {360 } (\bibinfo {year}
  {1973})}\BibitemShut {NoStop}%
\bibitem [{\citenamefont {Wang}\ and\ \citenamefont {Hioe}(1973)}]{wang73}%
  \BibitemOpen
  \bibfield  {author} {\bibinfo {author} {\bibfnamefont {Y.~K.}\ \bibnamefont
  {Wang}}\ and\ \bibinfo {author} {\bibfnamefont {F.~T.}\ \bibnamefont
  {Hioe}},\ }\bibfield  {title} {\enquote {\bibinfo {title} {Phase transition
  in the dicke model of superradiance},}\ }\href {\doibase
  10.1103/PhysRevA.7.831} {\bibfield  {journal} {\bibinfo  {journal} {Phys.
  Rev. A}\ }\textbf {\bibinfo {volume} {7}},\ \bibinfo {pages} {831} (\bibinfo
  {year} {1973})}\BibitemShut {NoStop}%
\bibitem [{\citenamefont {Dimer}\ \emph {et~al.}(2007)\citenamefont {Dimer},
  \citenamefont {Estienne}, \citenamefont {Parkins},\ and\ \citenamefont
  {Carmichael}}]{dimer07}%
  \BibitemOpen
  \bibfield  {author} {\bibinfo {author} {\bibfnamefont {F.}~\bibnamefont
  {Dimer}}, \bibinfo {author} {\bibfnamefont {B.}~\bibnamefont {Estienne}},
  \bibinfo {author} {\bibfnamefont {A.~S.}\ \bibnamefont {Parkins}}, \ and\
  \bibinfo {author} {\bibfnamefont {H.~J.}\ \bibnamefont {Carmichael}},\
  }\bibfield  {title} {\enquote {\bibinfo {title} {Proposed realization of the
  dicke-model quantum phase transition in an optical cavity qed system},}\
  }\href {\doibase 10.1103/PhysRevA.75.013804} {\bibfield  {journal} {\bibinfo
  {journal} {Phys. Rev. A}\ }\textbf {\bibinfo {volume} {75}},\ \bibinfo
  {pages} {013804} (\bibinfo {year} {2007})}\BibitemShut {NoStop}%
\bibitem [{\citenamefont {Baden}\ \emph {et~al.}(2014)\citenamefont {Baden},
  \citenamefont {Arnold}, \citenamefont {Grimsmo}, \citenamefont {Parkins},\
  and\ \citenamefont {Barrett}}]{baden14}%
  \BibitemOpen
  \bibfield  {author} {\bibinfo {author} {\bibfnamefont {M.~P.}\ \bibnamefont
  {Baden}}, \bibinfo {author} {\bibfnamefont {K.~J.}\ \bibnamefont {Arnold}},
  \bibinfo {author} {\bibfnamefont {A.~L.}\ \bibnamefont {Grimsmo}}, \bibinfo
  {author} {\bibfnamefont {S.}~\bibnamefont {Parkins}}, \ and\ \bibinfo
  {author} {\bibfnamefont {M.~D.}\ \bibnamefont {Barrett}},\ }\bibfield
  {title} {\enquote {\bibinfo {title} {Realization of the dicke model using
  cavity-assisted raman transitions},}\ }\href {\doibase
  10.1103/PhysRevLett.113.020408} {\bibfield  {journal} {\bibinfo  {journal}
  {Phys. Rev. Lett.}\ }\textbf {\bibinfo {volume} {113}},\ \bibinfo {pages}
  {020408} (\bibinfo {year} {2014})}\BibitemShut {NoStop}%
\bibitem [{\citenamefont {Black}\ \emph {et~al.}(2003)\citenamefont {Black},
  \citenamefont {Chan},\ and\ \citenamefont {Vuleti\ifmmode~\acute{c}\else
  \'{c}\fi{}}}]{black03}%
  \BibitemOpen
  \bibfield  {author} {\bibinfo {author} {\bibfnamefont {A.~T.}\ \bibnamefont
  {Black}}, \bibinfo {author} {\bibfnamefont {H.~W.}\ \bibnamefont {Chan}}, \
  and\ \bibinfo {author} {\bibfnamefont {V.}~\bibnamefont
  {Vuleti\ifmmode~\acute{c}\else \'{c}\fi{}}},\ }\bibfield  {title} {\enquote
  {\bibinfo {title} {Observation of collective friction forces due to spatial
  self-organization of atoms: From rayleigh to bragg scattering},}\ }\href
  {\doibase 10.1103/PhysRevLett.91.203001} {\bibfield  {journal} {\bibinfo
  {journal} {Phys. Rev. Lett.}\ }\textbf {\bibinfo {volume} {91}},\ \bibinfo
  {pages} {203001} (\bibinfo {year} {2003})}\BibitemShut {NoStop}%
\bibitem [{\citenamefont {Baumann}\ \emph {et~al.}(2010)\citenamefont
  {Baumann}, \citenamefont {Guerlin}, \citenamefont {Brennecke},\ and\
  \citenamefont {Esslinger}}]{baumann10}%
  \BibitemOpen
  \bibfield  {author} {\bibinfo {author} {\bibfnamefont {K.}~\bibnamefont
  {Baumann}}, \bibinfo {author} {\bibfnamefont {C.}~\bibnamefont {Guerlin}},
  \bibinfo {author} {\bibfnamefont {F.}~\bibnamefont {Brennecke}}, \ and\
  \bibinfo {author} {\bibfnamefont {T.}~\bibnamefont {Esslinger}},\ }\bibfield
  {title} {\enquote {\bibinfo {title} {Dicke quantum phase transition with a
  superfluid gas in an optical cavity},}\ }\href {\doibase
  doi:10.1038/nature09009} {\bibfield  {journal} {\bibinfo  {journal} {Nature}\
  }\textbf {\bibinfo {volume} {464}},\ \bibinfo {pages} {1301} (\bibinfo {year}
  {2010})}\BibitemShut {NoStop}%
\bibitem [{\citenamefont {Brennecke}\ \emph {et~al.}(2013)\citenamefont
  {Brennecke}, \citenamefont {Mottl}, \citenamefont {Baumann}, \citenamefont
  {Landig}, \citenamefont {Donner},\ and\ \citenamefont
  {Esslinger}}]{brennecke13}%
  \BibitemOpen
  \bibfield  {author} {\bibinfo {author} {\bibfnamefont {F.}~\bibnamefont
  {Brennecke}}, \bibinfo {author} {\bibfnamefont {R.}~\bibnamefont {Mottl}},
  \bibinfo {author} {\bibfnamefont {K.}~\bibnamefont {Baumann}}, \bibinfo
  {author} {\bibfnamefont {R.}~\bibnamefont {Landig}}, \bibinfo {author}
  {\bibfnamefont {T.}~\bibnamefont {Donner}}, \ and\ \bibinfo {author}
  {\bibfnamefont {T.}~\bibnamefont {Esslinger}},\ }\bibfield  {title} {\enquote
  {\bibinfo {title} {Real-time observation of fluctuations at the
  driven-dissipative dicke phase transition},}\ }\href {\doibase
  10.1073/pnas.1306993110} {\bibfield  {journal} {\bibinfo  {journal} {Proc.
  Natl. Acad. Sci. U.S.A.}\ }\textbf {\bibinfo {volume} {110}},\ \bibinfo
  {pages} {11763} (\bibinfo {year} {2013})}\BibitemShut {NoStop}%
\bibitem [{\citenamefont {Piazza}\ \emph {et~al.}(2013)\citenamefont {Piazza},
  \citenamefont {Strack},\ and\ \citenamefont {Zwerger}}]{piazza13}%
  \BibitemOpen
  \bibfield  {author} {\bibinfo {author} {\bibfnamefont {F.}~\bibnamefont
  {Piazza}}, \bibinfo {author} {\bibfnamefont {P.}~\bibnamefont {Strack}}, \
  and\ \bibinfo {author} {\bibfnamefont {W.}~\bibnamefont {Zwerger}},\
  }\bibfield  {title} {\enquote {\bibinfo {title} {Bose-einstein condensation
  versus dicke-hepp-lieb transition in an optical cavity},}\ }\href@noop {}
  {\bibfield  {journal} {\bibinfo  {journal} {Annals of Physics}\ }\textbf
  {\bibinfo {volume} {339}},\ \bibinfo {pages} {135} (\bibinfo {year}
  {2013})}\BibitemShut {NoStop}%
\bibitem [{\citenamefont {Kulkarni}\ \emph {et~al.}(2013)\citenamefont
  {Kulkarni}, \citenamefont {\"Oztop},\ and\ \citenamefont
  {T\"ureci}}]{kulkarni13}%
  \BibitemOpen
  \bibfield  {author} {\bibinfo {author} {\bibfnamefont {M.}~\bibnamefont
  {Kulkarni}}, \bibinfo {author} {\bibfnamefont {B.}~\bibnamefont {\"Oztop}}, \
  and\ \bibinfo {author} {\bibfnamefont {H.~E.}\ \bibnamefont {T\"ureci}},\
  }\bibfield  {title} {\enquote {\bibinfo {title} {Cavity-mediated
  near-critical dissipative dynamics of a driven condensate},}\ }\href
  {\doibase 10.1103/PhysRevLett.111.220408} {\bibfield  {journal} {\bibinfo
  {journal} {Phys. Rev. Lett.}\ }\textbf {\bibinfo {volume} {111}},\ \bibinfo
  {pages} {220408} (\bibinfo {year} {2013})}\BibitemShut {NoStop}%
\bibitem [{\citenamefont {K\'onya}\ \emph {et~al.}(2014)\citenamefont
  {K\'onya}, \citenamefont {Szirmai},\ and\ \citenamefont {Domokos}}]{konya14}%
  \BibitemOpen
  \bibfield  {author} {\bibinfo {author} {\bibfnamefont {G.}~\bibnamefont
  {K\'onya}}, \bibinfo {author} {\bibfnamefont {G.}~\bibnamefont {Szirmai}}, \
  and\ \bibinfo {author} {\bibfnamefont {P.}~\bibnamefont {Domokos}},\
  }\bibfield  {title} {\enquote {\bibinfo {title} {Damping of quasiparticles in
  a bose-einstein condensate coupled to an optical cavity},}\ }\href {\doibase
  10.1103/PhysRevA.90.013623} {\bibfield  {journal} {\bibinfo  {journal} {Phys.
  Rev. A}\ }\textbf {\bibinfo {volume} {90}},\ \bibinfo {pages} {013623}
  (\bibinfo {year} {2014})}\BibitemShut {NoStop}%
\bibitem [{\citenamefont {Piazza}\ and\ \citenamefont
  {Strack}(2014)}]{piazza14}%
  \BibitemOpen
  \bibfield  {author} {\bibinfo {author} {\bibfnamefont {F.}~\bibnamefont
  {Piazza}}\ and\ \bibinfo {author} {\bibfnamefont {P.}~\bibnamefont
  {Strack}},\ }\bibfield  {title} {\enquote {\bibinfo {title} {Quantum kinetics
  of ultracold fermions coupled to an optical resonator},}\ }\href {\doibase
  10.1103/PhysRevA.90.043823} {\bibfield  {journal} {\bibinfo  {journal} {Phys.
  Rev. A}\ }\textbf {\bibinfo {volume} {90}},\ \bibinfo {pages} {043823}
  (\bibinfo {year} {2014})}\BibitemShut {NoStop}%
\bibitem [{\citenamefont {Scully}(2015)}]{Scully2015}%
  \BibitemOpen
  \bibfield  {author} {\bibinfo {author} {\bibfnamefont {M.~O.}\ \bibnamefont
  {Scully}},\ }\bibfield  {title} {\enquote {\bibinfo {title} {Single photon
  subradiance: Quantum control of spontaneous emission and ultrafast
  readout},}\ }\href {\doibase 10.1103/PhysRevLett.115.243602} {\bibfield
  {journal} {\bibinfo  {journal} {Phys. Rev. Lett.}\ }\textbf {\bibinfo
  {volume} {115}},\ \bibinfo {pages} {243602} (\bibinfo {year}
  {2015})}\BibitemShut {NoStop}%
\bibitem [{\citenamefont {Thompson}\ \emph {et~al.}(2013)\citenamefont
  {Thompson}, \citenamefont {Tiecke}, \citenamefont {de~Leon}, \citenamefont
  {Feist}, \citenamefont {Akimov}, \citenamefont {Gullans}, \citenamefont
  {Zibrov}, \citenamefont {Vuleti{\'c}},\ and\ \citenamefont
  {Lukin}}]{thompson12}%
  \BibitemOpen
  \bibfield  {author} {\bibinfo {author} {\bibfnamefont {J.~D.}\ \bibnamefont
  {Thompson}}, \bibinfo {author} {\bibfnamefont {T.~G.}\ \bibnamefont
  {Tiecke}}, \bibinfo {author} {\bibfnamefont {N.~P.}\ \bibnamefont {de~Leon}},
  \bibinfo {author} {\bibfnamefont {J.}~\bibnamefont {Feist}}, \bibinfo
  {author} {\bibfnamefont {A.~V.}\ \bibnamefont {Akimov}}, \bibinfo {author}
  {\bibfnamefont {M.}~\bibnamefont {Gullans}}, \bibinfo {author} {\bibfnamefont
  {A.~S.}\ \bibnamefont {Zibrov}}, \bibinfo {author} {\bibfnamefont
  {V.}~\bibnamefont {Vuleti{\'c}}}, \ and\ \bibinfo {author} {\bibfnamefont
  {M.~D.}\ \bibnamefont {Lukin}},\ }\bibfield  {title} {\enquote {\bibinfo
  {title} {Coupling a single trapped atom to a nanoscale optical cavity},}\
  }\href {\doibase 10.1126/science.1237125} {\bibfield  {journal} {\bibinfo
  {journal} {Science}\ }\textbf {\bibinfo {volume} {340}},\ \bibinfo {pages}
  {1202} (\bibinfo {year} {2013})}\BibitemShut {NoStop}%
\bibitem [{\citenamefont {Chang}\ \emph {et~al.}(2013)\citenamefont {Chang},
  \citenamefont {Cirac},\ and\ \citenamefont {Kimble}}]{chang13}%
  \BibitemOpen
  \bibfield  {author} {\bibinfo {author} {\bibfnamefont {D.~E.}\ \bibnamefont
  {Chang}}, \bibinfo {author} {\bibfnamefont {J.~I.}\ \bibnamefont {Cirac}}, \
  and\ \bibinfo {author} {\bibfnamefont {H.~J.}\ \bibnamefont {Kimble}},\
  }\bibfield  {title} {\enquote {\bibinfo {title} {Self-organization of atoms
  along a nanophotonic waveguide},}\ }\href {\doibase
  10.1103/PhysRevLett.110.113606} {\bibfield  {journal} {\bibinfo  {journal}
  {Phys. Rev. Lett.}\ }\textbf {\bibinfo {volume} {110}},\ \bibinfo {pages}
  {113606} (\bibinfo {year} {2013})}\BibitemShut {NoStop}%
\bibitem [{\citenamefont {Dalla~Torre}\ \emph {et~al.}(2013)\citenamefont
  {Dalla~Torre}, \citenamefont {Diehl}, \citenamefont {Lukin}, \citenamefont
  {Sachdev},\ and\ \citenamefont {Strack}}]{dalla13}%
  \BibitemOpen
  \bibfield  {author} {\bibinfo {author} {\bibfnamefont {E.~G.}\ \bibnamefont
  {Dalla~Torre}}, \bibinfo {author} {\bibfnamefont {S.}~\bibnamefont {Diehl}},
  \bibinfo {author} {\bibfnamefont {M.~D.}\ \bibnamefont {Lukin}}, \bibinfo
  {author} {\bibfnamefont {S.}~\bibnamefont {Sachdev}}, \ and\ \bibinfo
  {author} {\bibfnamefont {P.}~\bibnamefont {Strack}},\ }\bibfield  {title}
  {\enquote {\bibinfo {title} {Keldysh approach for nonequilibrium phase
  transitions in quantum optics: Beyond the dicke model in optical cavities},}\
  }\href {\doibase 10.1103/PhysRevA.87.023831} {\bibfield  {journal} {\bibinfo
  {journal} {Phys. Rev. A}\ }\textbf {\bibinfo {volume} {87}},\ \bibinfo
  {pages} {023831} (\bibinfo {year} {2013})}\BibitemShut {NoStop}%
\bibitem [{\citenamefont {Habibian}\ \emph {et~al.}(2013)\citenamefont
  {Habibian}, \citenamefont {Winter}, \citenamefont {Paganelli}, \citenamefont
  {Rieger},\ and\ \citenamefont {Morigi}}]{habibian13}%
  \BibitemOpen
  \bibfield  {author} {\bibinfo {author} {\bibfnamefont {H.}~\bibnamefont
  {Habibian}}, \bibinfo {author} {\bibfnamefont {A.}~\bibnamefont {Winter}},
  \bibinfo {author} {\bibfnamefont {S.}~\bibnamefont {Paganelli}}, \bibinfo
  {author} {\bibfnamefont {H.}~\bibnamefont {Rieger}}, \ and\ \bibinfo {author}
  {\bibfnamefont {G.}~\bibnamefont {Morigi}},\ }\bibfield  {title} {\enquote
  {\bibinfo {title} {Bose-glass phases of ultracold atoms due to cavity
  backaction},}\ }\href {\doibase 10.1103/PhysRevLett.110.075304} {\bibfield
  {journal} {\bibinfo  {journal} {Phys. Rev. Lett.}\ }\textbf {\bibinfo
  {volume} {110}},\ \bibinfo {pages} {075304} (\bibinfo {year}
  {2013})}\BibitemShut {NoStop}%
\bibitem [{\citenamefont {Bhaseen}\ \emph {et~al.}(2012)\citenamefont
  {Bhaseen}, \citenamefont {Mayoh}, \citenamefont {Simons},\ and\ \citenamefont
  {Keeling}}]{bhaseen12}%
  \BibitemOpen
  \bibfield  {author} {\bibinfo {author} {\bibfnamefont {M.~J.}\ \bibnamefont
  {Bhaseen}}, \bibinfo {author} {\bibfnamefont {J.}~\bibnamefont {Mayoh}},
  \bibinfo {author} {\bibfnamefont {B.~D.}\ \bibnamefont {Simons}}, \ and\
  \bibinfo {author} {\bibfnamefont {J.}~\bibnamefont {Keeling}},\ }\bibfield
  {title} {\enquote {\bibinfo {title} {Dynamics of nonequilibrium dicke
  models},}\ }\href {\doibase 10.1103/PhysRevA.85.013817} {\bibfield  {journal}
  {\bibinfo  {journal} {Phys. Rev. A}\ }\textbf {\bibinfo {volume} {85}},\
  \bibinfo {pages} {013817} (\bibinfo {year} {2012})}\BibitemShut {NoStop}%
\bibitem [{Note1()}]{Note1}%
  \BibitemOpen
  \bibinfo {note} {M. Barrett/Singapore Experiment, Private
  Communication}\BibitemShut {NoStop}%
\bibitem [{\citenamefont {Keeling}\ \emph {et~al.}(2010)\citenamefont
  {Keeling}, \citenamefont {Bhaseen},\ and\ \citenamefont
  {Simons}}]{Keeling10}%
  \BibitemOpen
  \bibfield  {author} {\bibinfo {author} {\bibfnamefont {J.}~\bibnamefont
  {Keeling}}, \bibinfo {author} {\bibfnamefont {M.~J.}\ \bibnamefont
  {Bhaseen}}, \ and\ \bibinfo {author} {\bibfnamefont {B.~D.}\ \bibnamefont
  {Simons}},\ }\bibfield  {title} {\enquote {\bibinfo {title} {Collective
  dynamics of bose-einstein condensates in optical cavities},}\ }\href
  {\doibase 10.1103/PhysRevLett.105.043001} {\bibfield  {journal} {\bibinfo
  {journal} {Phys. Rev. Lett.}\ }\textbf {\bibinfo {volume} {105}},\ \bibinfo
  {pages} {043001} (\bibinfo {year} {2010})}\BibitemShut {NoStop}%
\bibitem [{\citenamefont {Nagy}\ \emph {et~al.}(2008)\citenamefont {Nagy},
  \citenamefont {Szirmai},\ and\ \citenamefont {Domokos}}]{Nagy2008}%
  \BibitemOpen
  \bibfield  {author} {\bibinfo {author} {\bibfnamefont {D.}~\bibnamefont
  {Nagy}}, \bibinfo {author} {\bibfnamefont {G.}~\bibnamefont {Szirmai}}, \
  and\ \bibinfo {author} {\bibfnamefont {P.}~\bibnamefont {Domokos}},\
  }\bibfield  {title} {\enquote {\bibinfo {title} {Self-organization of a
  bose-einstein condensate in an optical cavity},}\ }\href {\doibase
  10.1140/epjd/e2008-00074-6} {\bibfield  {journal} {\bibinfo  {journal} {The
  European Physical Journal D}\ }\textbf {\bibinfo {volume} {48}},\ \bibinfo
  {pages} {127} (\bibinfo {year} {2008})}\BibitemShut {NoStop}%
\bibitem [{\citenamefont {Collett}\ and\ \citenamefont
  {Gardiner}(1984)}]{collett84}%
  \BibitemOpen
  \bibfield  {author} {\bibinfo {author} {\bibfnamefont {M.~J.}\ \bibnamefont
  {Collett}}\ and\ \bibinfo {author} {\bibfnamefont {C.~W.}\ \bibnamefont
  {Gardiner}},\ }\bibfield  {title} {\enquote {\bibinfo {title} {Squeezing of
  intracavity and traveling-wave light fields produced in parametric
  amplification},}\ }\href {\doibase 10.1103/PhysRevA.30.1386} {\bibfield
  {journal} {\bibinfo  {journal} {Phys. Rev. A}\ }\textbf {\bibinfo {volume}
  {30}},\ \bibinfo {pages} {1386} (\bibinfo {year} {1984})}\BibitemShut
  {NoStop}%
\bibitem [{\citenamefont {Gardiner}\ and\ \citenamefont
  {Collett}(1985)}]{gardiner85}%
  \BibitemOpen
  \bibfield  {author} {\bibinfo {author} {\bibfnamefont {C.~W.}\ \bibnamefont
  {Gardiner}}\ and\ \bibinfo {author} {\bibfnamefont {M.~J.}\ \bibnamefont
  {Collett}},\ }\bibfield  {title} {\enquote {\bibinfo {title} {Input and
  output in damped quantum systems: Quantum stochastic differential equations
  and the master equation},}\ }\href {\doibase 10.1103/PhysRevA.31.3761}
  {\bibfield  {journal} {\bibinfo  {journal} {Phys. Rev. A}\ }\textbf {\bibinfo
  {volume} {31}},\ \bibinfo {pages} {3761} (\bibinfo {year}
  {1985})}\BibitemShut {NoStop}%
\bibitem [{\citenamefont {Zeiher}\ \emph {et~al.}(2016)\citenamefont {Zeiher},
  \citenamefont {van Biijnen}, \citenamefont {Schauss}, \citenamefont {Hild},
  \citenamefont {Choi}, \citenamefont {Pohl}, \citenamefont {Bloch},\ and\
  \citenamefont {Gross}}]{zeiher16}%
  \BibitemOpen
  \bibfield  {author} {\bibinfo {author} {\bibfnamefont {J.}~\bibnamefont
  {Zeiher}}, \bibinfo {author} {\bibfnamefont {R.}~\bibnamefont {van Biijnen}},
  \bibinfo {author} {\bibfnamefont {P.}~\bibnamefont {Schauss}}, \bibinfo
  {author} {\bibfnamefont {S.}~\bibnamefont {Hild}}, \bibinfo {author}
  {\bibfnamefont {J.}~\bibnamefont {Choi}}, \bibinfo {author} {\bibfnamefont
  {T.}~\bibnamefont {Pohl}}, \bibinfo {author} {\bibfnamefont {I.}~\bibnamefont
  {Bloch}}, \ and\ \bibinfo {author} {\bibfnamefont {C.}~\bibnamefont
  {Gross}},\ }\bibfield  {title} {\enquote {\bibinfo {title} {Many-body
  interferometry of a rydberg-dressed spin lattice},}\ }\href {\doibase
  doi:10.1038/nphys3835} {\bibfield  {journal} {\bibinfo  {journal} {Nat.
  Phys.}\ }\textbf {\bibinfo {volume} {12}},\ \bibinfo {pages} {1095} (\bibinfo
  {year} {2016})}\BibitemShut {NoStop}%
\bibitem [{\citenamefont {Scully}\ and\ \citenamefont
  {Zubairy}(1997)}]{scully}%
  \BibitemOpen
  \bibfield  {author} {\bibinfo {author} {\bibfnamefont {M.}~\bibnamefont
  {Scully}}\ and\ \bibinfo {author} {\bibfnamefont {M.}~\bibnamefont
  {Zubairy}},\ }\href {https://books.google.co.uk/books?id=20ISsQCKKmQC} {\emph
  {\bibinfo {title} {Quantum Optics}}},\ \bibinfo {edition} {1st}\ ed.\
  (\bibinfo  {publisher} {Cambridge University Press},\ \bibinfo {year}
  {1997})\BibitemShut {NoStop}%
\bibitem [{\citenamefont {Steck}(2007)}]{Steck}%
  \BibitemOpen
  \bibfield  {author} {\bibinfo {author} {\bibfnamefont {D.~A.}\ \bibnamefont
  {Steck}},\ }\href {http://steck.us/teaching} {\emph {\bibinfo {title}
  {Quantum and Atom Optics}}},\ \bibinfo {edition} {1st}\ ed.\ (\bibinfo {year}
  {2007})\BibitemShut {NoStop}%
\bibitem [{\citenamefont {Arnold}\ \emph {et~al.}(2012)\citenamefont {Arnold},
  \citenamefont {Baden},\ and\ \citenamefont {Barrett}}]{Barrett2012}%
  \BibitemOpen
  \bibfield  {author} {\bibinfo {author} {\bibfnamefont {K.~J.}\ \bibnamefont
  {Arnold}}, \bibinfo {author} {\bibfnamefont {M.~P.}\ \bibnamefont {Baden}}, \
  and\ \bibinfo {author} {\bibfnamefont {M.~D.}\ \bibnamefont {Barrett}},\
  }\bibfield  {title} {\enquote {\bibinfo {title} {Self-organization threshold
  scaling for thermal atoms coupled to a cavity},}\ }\href {\doibase
  10.1103/PhysRevLett.109.153002} {\bibfield  {journal} {\bibinfo  {journal}
  {Phys. Rev. Lett.}\ }\textbf {\bibinfo {volume} {109}},\ \bibinfo {pages}
  {153002} (\bibinfo {year} {2012})}\BibitemShut {NoStop}%
\end{thebibliography}%

\end{document}